\documentclass{tlp}
\usepackage{aopmath}
\usepackage{amssymb}
\usepackage{graphicx}

\usepackage[all, knot]{xy} 
\input xy
\xyoption{all}

\newcommand{\arat}{\ar@}

\newtheorem{definition}{Definition}[section]

\begin{document}

\bibliographystyle{acmtrans}






\newcommand{\bi}{\begin{array}[t]{@{}l@{}}}
\newcommand{\ei}{\end{array}}
\newcommand{\ba}{\begin{array}}
\newcommand{\ea}{\end{array}}
\newcommand{\bda}{\[\ba}
\newcommand{\eda}{\ea\]}
\newcommand{\bp}{\begin{quote}\tt\begin{tabbing}}
\newcommand{\ep}{\end{tabbing}\end{quote}}


\newcommand{\good}{good}
 
\newcommand{\W}{\mbox{${\cal W}$}}

\newcommand{\err}{{\bf W}}
\newcommand{\VV}{{\cal V}} 
\newcommand{\GT}{{\cal T}_G}
\newcommand{\KK}{{\cal K}}
\newcommand{\TT}{{\cal T}}
\newcommand{\TC}{\mbox{\it TC}}
\newcommand{\UL}{\mbox{UL}}
\newcommand{\sUL}{\mbox{\scriptsize UL}}
\newcommand{\true}{\mbox{\sf true}}
\newcommand{\Int}{\mbox{\it Int}}
\newcommand{\Bool}{\mbox{\it Bool}}
\newcommand{\SF}{\mbox{$\cal S$}}
\newcommand{\tauvec}{\bar{\tau}}
\newcommand{\tvec}{\bar{t}}
\newcommand{\kvec}{\bar{k}}
\newcommand{\xvec}{\bar{x}}
\newcommand{\muvec}{\bar{\mu}}
\newcommand{\alphavec}{\bar{\alpha}}
\newcommand{\betavec}{\bar{\beta}}
\newcommand{\gammavec}{\bar{\gamma}}
\newcommand{\thetavec}{\bar{\theta}}
\newcommand{\deltavec}{\bar{\delta}}
\newcommand{\bvec}{\bar{b}}
\newcommand{\typo}{\Gamma}
\newcommand{\tenv}{\Gamma}
\newcommand{\typoinit}{\typo_0}
\newcommand{\static}{S}
\newcommand{\dynamic}{D}
\newcommand{\sstatic}{{\scriptstyle S}}
\newcommand{\sdynamic}{{\scriptstyle D}}
\newcommand{\BTA}{\mbox{BTA}}
\newcommand{\baseshape}{\bot}
\newcommand{\TCT}{P}
\newcommand{\Haskell}{H98}
\newcommand{\TCTinit}{P_o}


\newcommand{\tlabel}[1]{\mbox{(#1)}}
\newcommand{\rlabel}[1]{\mbox{\it (#1)}}
\newcommand{\fig}[3]
        {\begin{figure*}[t]#3\
        \caption{\label{#1}#2}\ \hrulefill\ \end{figure*}}
\newcommand{\figurebox}[1]
        {\fbox{\begin{minipage}{\textwidth} #1 \end{minipage}}}
\newcommand{\boxfig}[3]
        {\begin{figure*}\figurebox{#3\caption{\label{#1}#2}}\end{figure*}}
\newcommand{\myirule}[2]{{\renewcommand{\arraystretch}{1.2}\ba{c} #1
                      \\ \hline #2 \ea}}


\newcommand{\llist}[1]{\langle #1 \rangle}
\newcommand{\sat}{\mbox{\it sat}}
\newcommand{\entail}{\mbox{\it entail}}
\newcommand{\unambig}{\mbox{\it unambig}}
\newcommand{\complete}{\mbox{\it complete}}

\newcommand{\projexclusion}[1]{\mbox{$\bar{\exists}#1$}}
\newcommand{\funinst}{\mbox{\it inst}}
\newcommand{\ileq}{\preceq}
\newcommand{\ieq}{\simeq}
\newcommand{\topannot}[1]{|#1|}
\newcommand{\gendelta}{\mbox{\it gen}_{\delta}}
\newcommand{\genbeta}{\mbox{\it gen}_{\beta}}
\newcommand{\gen}{\mbox{\it gen}}
\newcommand{\normalize}{\mbox{\it normalize}}
\newcommand{\fail}{\mbox{\it fail}}
\newcommand{\fixpt}{\mbox{${\cal F}$}}
\newcommand{\testpa}{\mbox{${\cal T}$}}
\newcommand{\addpa}{\mbox{${\cal A}$}}
\newcommand{\matchpa}{\mbox{${\cal M}$}}
\newcommand{\refmatchpa}{\mbox{${\cal R}$}}
\newcommand{\wft}[1]{\mbox{\it wft}(#1)}
\newcommand{\tv}{\mbox{\it fv}}
\newcommand{\stv}{\mbox{\it \scriptsize fv}}
\newcommand{\range}{\mbox{\it range}}
\newcommand{\lub}{\wedge}
\newcommand{\tyrel}[2]{(#1 \preceq #2)}
\newcommand{\eqty}[2]{#1=#2}
\newcommand{\turns}{\, \vdash \,}
\newcommand{\tsinst}{\, \vdash^i \,}
\newcommand{\tsttv}{\, \vdash^{ttv} \,}
\newcommand{\tsUL}{\, \vdash_{\mbox{\scriptsize UL}} \,}
\newcommand{\trc}{\, \vdash_{\scriptsize \id{inf}} \,}
\newcommand{\trl}{\, \vdash_{\scriptsize \id{W}} \,}
\newcommand{\trcp}{\, \vdash_{\scriptstyle inf}' \,}
\newcommand{\genrel}{\, \vdash^{\scriptstyle gen} \,}
\newcommand{\asubty}[2]{(#1 \leq_a #2)}
\newcommand{\ssubty}[2]{(#1 \leq_s #2)}
\newcommand{\fsubty}[2]{(#1 \leq_f #2)}
\newcommand{\subty}[2]{(#1 \leq #2)}
\newcommand{\doubleb}[1]{[\![ #1 ]\!]}

\newcommand{\mynote}[1]{$\spadesuit${\bf #1}$\clubsuit$}


\newcommand{\sgap}{\quad}
\newcommand{\bgap}{\quad\quad}


\newcommand{\mathem}{\sf}
\newcommand{\IN}{\mbox{\mathem in}}
\newcommand{\LET}{\mbox{\mathem let}}
\newcommand{\MLET}{\mbox{\mathem mlet}}
\newcommand{\LETREC}{\mbox{\mathem letrec}}
\newcommand{\FIX}{\mbox{\mathem fix}}
\newcommand{\CASE}{\mbox{\mathem case}}
\newcommand{\TYCASE}{\mbox{\mathem typecase}}
\newcommand{\ELSE}{\mbox{\mathem else}}
\newcommand{\IF}{\mbox{\mathem if}}
\newcommand{\OF}{\mbox{\mathem of}}
\newcommand{\THEN}{\mbox{\mathem then}}
\newcommand{\INST}{\mbox{\mathem instance}}
\newcommand{\OVER}{\mbox{\mathem overload}}
\newcommand{\CLASS}{\mbox{\mathem class}}
\newcommand{\WHERE}{\mbox{\mathem where}}


\def\ruleform#1{{\setlength{\fboxrule}{1pt}\fbox{\normalsize $#1$}}}
\newcommand{\ms}[1]{\marginpar{\sc ms}{\bf #1}}

\newcommand{\atsign}{@}

\newcommand{\simparrow}[0]{\Longleftrightarrow}
\newcommand{\proparrow}[0]{\Longrightarrow}
\newcommand{\arrow}[0]{\rightarrow}

\renewcommand{\rlabel}[1]{\mbox{$\mathit{#1}$}}
\newcommand{\blabel}[1]{\mbox{\bf (#1)}}
\newcommand{\Leq}{\mathit{Leq}}
\newcommand{\LL}{\mathit{L}}
\newcommand{\Merge}{\mathit{Merge}}
\newcommand{\MM}{\mathit{M}}
\newcommand{\Get}{\mathit{Get}}
\newcommand{\Put}{\mathit{Put}}
\newcommand{\Gcd}{\mathit{Gcd}}
\newcommand{\True}{\mathit{True}}

\newcommand{\ignore}[1]{}

\newcommand{\abssem}{{\cal A}}
\newcommand{\goalsem}{{\cal G}}
\newcommand{\pargoalsem}{{\parallel \cal G}}

\newcommand{\abstrans}{\rightarrowtail_{\cal A}}
\newcommand{\abstransstar}{\rightarrowtail^*_{\cal A}}
\newcommand{\goaltrans}{\rightarrowtail_{\cal G}}
\newcommand{\goaltransstar}{\rightarrowtail^*_{\cal G}}
\newcommand{\partrans}{\rightarrowtail_{\mid\mid\cal G}}
\newcommand{\partransstar}{\rightarrowtail^*_{\mid\mid\cal G}}

\newcommand{\abscup}{\uplus}
\newcommand{\goalcup}{\uplus}
\newcommand{\stcup}{\cup}

\newcommand{\goaltranssf}[1]{\stackrel{#1}{\rightarrowtail_{\cal G}}}
\newcommand{\partranssf}[1]{\stackrel{#1}{\rightarrowtail_{\mid\mid\cal G}}}

\newcommand{\chrstate}[2]{\langle #1\mid#2 \rangle}

\long\def\comment#1{}

\title{Concurrent Goal-Based Execution of Constraint Handling Rules}

\author[Edmund S. L. Lam and Martin Sulzmann]{
Edmund S. L. Lam \\
 School of Computing, National University of Singapore \\
 S16 Level 5, 3 Science Drive 2, Singapore 117543 \\
 {\tt lamsoonl@comp.nus.edu.sg} 
 \and 
Martin Sulzmann \\      
 Informatik Consulting Systems AG  \\
 {\tt martin.sulzmann@gmail.com}
}

\pagerange{\pageref{firstpage}--\pageref{lastpage}}
\setcounter{page}{1}

\maketitle

\label{firstpage}

\begin{abstract}
We introduce a systematic, concurrent execution scheme for 
Constraint Handling Rules (CHR) based on a previously proposed
sequential goal-based CHR semantics. We establish strong correspondence results to 
the abstract CHR semantics, thus guaranteeing that any answer in the 
concurrent, goal-based CHR semantics is reproducible
in the abstract CHR semantics.
Our work provides the foundation to obtain efficient, parallel CHR execution
schemes.
\end{abstract}


\section{Introduction}

Rewriting is a powerful discipline to specify the semantics of programming languages
and to perform automated deduction. There are numerous flavors of rewriting
such as term, graph rewriting etc. Our focus here is on
exhaustive, forward chaining, multi-set constraint rewriting as 
found in Constraint Handling Rules (CHR)~\cite{fruehwirth98:chr:art}
which are used in a multitude of applications such as
general purpose constraint programming,
type system design, agent specification and planning etc~\cite{1140337}.
Rewriting steps are specified via CHR rules which replace
a multi-set of constraints matching the left-hand side of a rule (also known as rule head) by
the rule's right-hand side (also known as rule body).

CHR support a very fine-grained form of concurrency.
CHR rules can be applied concurrently
if the rewriting steps they imply do not interfere with each other.
An interesting feature of CHR is that
the left-hand side of  a CHR rule
can have a mix of simplified and propagated constraint patterns.
This provides the opportunity for further concurrency. We can execute CHR rules
concurrently as long as only their propagated parts overlap.

The fact that the abstract CHR semantics is highly concurrent 
has so far not been exploited in any major CHR implementation.
Existing implementations are specified by 
highly deterministic semantics~\cite{DuckSBH04,rp-chr} which
support efficient and systematic but inherently 
single-threaded execution schemes~\cite{greg:thesis,DBLP:conf/iclp/Schrijvers05}.
Our goal is to develop a systematic, yet concurrent, semantics which can
be efficiently executed in parallel on a multi-core architecture.
In the CHR context, there is practically no prior work which addresses this important
issue.

Specifically, we make the following contributions:
\begin{itemize}
 \item We develop a novel goal-based concurrent CHR semantics.
 \item We verify that our semantics respects
        the abstract CHR semantics by establishing precise
        correspondence results.
 \item We examine which existing sequential CHR optimizations carry over
       to the concurrent setting.
\end{itemize}
Section~\ref{sec:concurrent-chr} contains the details.
A concrete parallel implementation derived from our concurrent semantics is
studied elsewhere~\cite{parallel-chr}. 
Section~\ref{sec:from-concurrent-to-parallel} provides a summary.

The upcoming section gives an overview of our work.
Section~\ref{sec:chr-example} reviews the abstract CHR semantics.
Section~\ref{sec:related} discusses prior work on execution schemes 
for CHR and production rule systems which are a related rewriting mechanism.
Section~\ref{sec:conc} concludes.

\section{Overview}

We first motivate concurrent execution of CHR rules via a few examples.
Then, we review existing deterministic CHR execution schemes
which are the basis for our concurrent goal-based CHR semantics.

\fig{f:chr-examples}{Communication channel and greatest common divisor}{

\mbox{Communication channel:}
\bda{c}
   \mathit{get} ~\atsign~ \Get(x),\Put(y) \simparrow x=y
\\ \\
\myirule{
    \{\Get(m),\Put(1)\} \rightarrowtail_{get} \{ m=1 \} 
   \parallel
  \{\Get(n),\Put(8)\}  \rightarrowtail_{get} \{ n=8 \}
  }
  {\{\Get(m),\Put(1),\Get(n),\Put(8)\} \rightarrowtail^* 
      \{ m=1, n=8 \}
  }
\eda

\vspace{4mm}

\mbox{Greatest common divisor:}
\bda{c}
  \mathit{gcd1}~\atsign~\Gcd(0)  \simparrow   \True
\\  \mathit{gcd2}~\atsign~\Gcd(n) \backslash \Gcd(m)  \simparrow   m>=n \&\& n>0 \mid \Gcd(m-n)
\\ \\
\myirule{
 \myirule{
   \ba{ll}
       & \{ \Gcd(3),\Gcd(9) \}
        \rightarrowtail_{gcd2}  \{\Gcd(3),\Gcd(6)  \} 
\\   \parallel
\\ &     \{ \Gcd(3), \Gcd(3) \}
       \rightarrowtail_{gcd2} \{ \Gcd(3), \Gcd(0) \}
 \ea
   }
  {  \ba{lll} \{\Gcd(3),\Gcd(3),\Gcd(9) \} & 
        \rightarrowtail_{gcd2,gcd2}  & \{ \Gcd(3),\Gcd(0),\Gcd(6) \} 
     \\ & \rightarrowtail^* & \{ \Gcd(3) \}
     \ea}
 }
 { \{\Gcd(3),\Gcd(3),\Gcd(9) \} \rightarrowtail^* \{ \Gcd(3) \} }
\eda
}

\fig{f:merge-sort}{Merge sort}{

\bda{c}
 merge1~\atsign~\Leq(x,a) \backslash \Leq(x,b) \simparrow a<b \mid \Leq(a,b)
\\ merge2~\atsign~\Merge(n,a),\Merge(n,b) \simparrow a<b \mid \Leq(a,b),\Merge(n+1,a)
\\ \\
\mbox{Shorthands: $\LL = \Leq$ and $\MM = \Merge$}
\\ \\
\myirule{
\myirule{
\ba{ll}
 &
 \ba{ll}
   & \MM(1,a), \MM(1,c), \MM(1,e), \MM(1,g)
 \\ \rightarrowtail_{merge2} & \MM(2,a),\MM(1,c), \MM(1,e), \LL(a,g)
 \\ \rightarrowtail_{merge2} & \MM(2,a), \MM(2,c), \LL(a,g), \LL(c,e)
 \\ \rightarrowtail_{merge2} & \MM(3,a), \LL(a,g), \LL(c,e), \LL(a,c)
\\ \rightarrowtail_{merge1} & \MM(3,a), \LL(a,c), \LL(c,g), \LL(c,e)
\\ \rightarrowtail_{merge1} & \MM(3,a), \LL(a,c), \LL(c,e), \LL(e,g)
 \ea
 \\ \\
 \parallel &
\\ \\
 &
 \ba{ll}
   & \MM(1,b), \MM(1,d), \MM(1,f), \MM(1,h)
 \\ \rightarrowtail^* & \MM(3,b), \LL(b,d), \LL(d,f), \LL(f,h)
 \ea
\ea 
}
{
\ba{ll}
 & \MM(3,a), \LL(a,c), \LL(c,e), \LL(e,g), \MM(3,b), \LL(b,d), \LL(d,f), \LL(f,h)
\\ \rightarrowtail_{merge2} & \MM(4,a), \LL(a,c),  \LL(a,b), \LL(c,e), \LL(e,g), \LL(b,d), \LL(d,f), \LL(f,h)
\\ \rightarrowtail_{merge1} & \MM(4,a), \LL(a,b),  \LL(b,c),\LL(c,e), \LL(e,g), \LL(b,d), \LL(d,f), \LL(f,h)
\\ \rightarrowtail_{merge1} & \MM(4,a), \LL(a,b),  \LL(b,c), \LL(c,d), \LL(c,e), \LL(e,g), \LL(d,f), \LL(f,h)
\\ \rightarrowtail_{merge1} & \MM(4,a), \LL(a,b),  \LL(b,c), \LL(c,d), \LL(d,e), \LL(e,g), \LL(d,f), \LL(f,h)
\\ \rightarrowtail_{merge1} & \MM(4,a), \LL(a,b),  \LL(b,c), \LL(c,d), \LL(d,e), \LL(e,f), \LL(e,g), \LL(f,h)
\\ \rightarrowtail_{merge1} & \MM(4,a), \LL(a,b),  \LL(b,c), \LL(c,d), \LL(d,e), \LL(e,f), \LL(f,g), \LL(f,h)
\\ \rightarrowtail_{merge1} & \MM(4,a), \LL(a,b),  \LL(b,c), \LL(c,d), \LL(d,e), \LL(e,f), \LL(f,g), \LL(g,h)
\ea
}
}
{
\ba{ll}
 & \MM(1,a), \MM(1,c), \MM(1,e), \MM(1,g),\MM(1,b), \MM(1,d), \MM(1,f), \MM(1,h)
\\ \rightarrowtail^* & \MM(4,a), \LL(a,b),  \LL(b,c), \LL(c,d), \LL(d,e), \LL(e,f), \LL(f,g), \LL(g,h)
\ea
}
\eda
}

\subsection{CHR and Concurrency}

Figures~\ref{f:chr-examples} 
and~\ref{f:merge-sort} contain several examples of CHR rules and derivations.
We adopt the convention that lower-case symbols refer to variables and upper-case symbols refer to
constraints. The notation $\mathit{rulename \atsign}$ assigns distinct labels to CHR rules.

The first example simulates a simple communication channel.
The $\mathit{Get(x)}$ constraint represents the action of writing a value from the communication
channel into the variable $x$, while the $\mathit{Put(y)}$ constraint represents the action
of putting the value $y$ into the channel.
The interaction between both constraints is specified via the CHR rule \rlabel{get} which
specifies the  replacement of constraints matching $\mathit{Get(x)}$ and $\mathit{Put(y)}$ by $x=y$.
The point to note is that
in contrast to Prolog, we use matching and not unification to trigger rules.

For example, the constraint store $\mathit{\{Get(m),Put(1)\}}$ matches the left-hand side of 
the \rlabel{get} rule
by instantiating $x$ by $m$ and $y$ by $1$. Hence, 
$\{\Get(m),\Put(1)\}$
rewrites to the answer $\{m=1\}$.
We write $\{\Get(m),\Put(1)\} \rightarrowtail_{get} \{ m=1 \}$ to denote this derivation step.
Similarly, we find that
$\{\Get(n),\Put(8)\} \rightarrowtail_{get} \{ n=8 \}$.
Rules can be applied concurrently as long as they do not interfere.
In our case, the two derivations above can be concurrently executed, indicated by the 
symbol $\parallel$, and we can straightforwardly combine both derivations
which leads to the final answer $\{ m=1, n=8 \}$.
We write $\rightarrowtail^*$ to denote exhaustive rule application.

The answer $\{m=8,n=1\}$ is also possible but the CHR rewrite semantics is committed-choice. 
We can guarantee a unique answer if the CHR rules are confluent which
means that rewritings applicable on overlapping constraint sets are always joinable.
In general, (non)confluence is of no concern to us here and is left to the programmer (if desired).
We follow here the abstract CHR semantics~\cite{fruehwirth98:chr:art}
(formally defined in Section~\ref{sec:chr-example})
which is inherently indeterministic.  Rewrite rules can be applied in any order
and thus CHR enjoy a high degree of concurrency.

The key to concurrency in CHR is monotonicity which 
guarantees that CHR executions remain valid if we include a larger context (i.e.~store).
The following result has been formally verified in~\cite{Abdennadher99confluenceand},
\begin{theorem} [Monotonicity of CHR] \label{theo:monotonicity}
    For any sets of CHR constraints $A$,$B$ and $S$, if 
   $A \rightarrowtail^* B$ then $A \uplus S \rightarrowtail^* B \uplus S$
\end{theorem}
An immediate consequence of monotonicity is that concurrent CHR executions are sound
in the sense that their effect can be reproduced using an appropriate sequential
sequence of execution steps. Thus, we can derive the following  rule:
\bda{cc}
\tlabel{\bf Concurrency} &
  \myirule{S\uplus S_1 \rightarrowtail^* S \uplus S_2 \sgap 
           S\uplus S_3 \rightarrowtail^* S \uplus S_4}
          {S\uplus S_1 \uplus S_3 \rightarrowtail^* S \uplus S_2 \uplus S_4}
\eda

In~\cite{union-find}, the above is referred to as "Strong Parallelism of CHR".
However, we prefer to use the term "concurrency" instead of "parallelism".
In the CHR context, concurrency means to run a CHR program (i.e.~a set of CHR rules)
by using concurrent execution threads.

Let's consider the second CHR example from Figure~\ref{f:chr-examples}  
which computes the greatest common divisor among a set of numbers
by applying Euclid's algorithm. 
The left-hand side of rule \rlabel{gcd2} is interesting
because it uses a mix of simplified and propagated constraint patterns.
We replace (simplify) $\mathit{Gcd(m)}$ by $\mathit{Gcd(m-n)}$ but keep (propagate) $\mathit{Gcd(n)}$
if the guard $m>=n \&\& n>0$ holds.
For example, we find that
$\{ \mathit{Gcd(3),Gcd(9)} \}
\rightarrowtail_{gcd2}  \{\mathit{Gcd(3),Gcd(6)}  \}$
and  $\{ \mathit{Gcd(3), Gcd(3)} \}
\rightarrowtail_{gcd2} \{ \mathit{Gcd(3), Gcd(0)} \}$.
The point to note is the above rule applications only overlap on the propagated part.
Hence, we can execute both rewrite derivations simultaneously
$$\{ \mathit{Gcd(3),Gcd(3) Gcd(9)} \} \rightarrowtail_{2 \times gcd2} \{ \mathit{Gcd(3), Gcd(0), Gcd(6)} \}$$

Our last example in Figure~\ref{f:merge-sort} is a CHR encoding of the well-known merge sort algorithm.
To sort a sequence of (distinct) elements $e_1,...,e_m$ where
$m$ is a power of $2$, we apply the rules to
the initial constraint store 
$$\mathit{\Merge(1,e_1),...,\Merge(1,e_m)}$$
Constraint $\mathit{\Merge(n,e)}$ refers to a sorted sequence of numbers at level $n$ whose smallest
element is $e$. Constraint $\Leq(a,b)$ denotes that $a$ is less than $b$.
Rule \rlabel{merge2} initiates the merging of two sorted lists and creates
a new sorted list at the next level. The actual merging
is performed by rule \rlabel{merge1}.
Sorting of sublists belonging to different mergers can be performed simultaneously.
See the example derivation in Figure~\ref{f:merge-sort} where
we simultaneously sort the characters $a,c,e,g$ and $b,d,f,h$.

\subsection{Goal-Based CHR Execution}

Existing CHR implementation employ a more systematic
CHR execution model where rules are triggered based on a set of available goals.
The idea behind a goal-based CHR execution model is  
to separate the constraint store  into
two components: a set of goal constraints (constraints yet to be executed)
and the actual constraint store (constraints that were executed). 
Previously, in the abstract semantics transitions $\abstrans$ are among
states $\mathit{Store}$ whereas in the goal-based semantics we find now transitions
$\goaltrans$ among states of the form $\chrstate{\mathit{Goals}}{\mathit{Store}}$.
Only goal constraints can trigger rules by 
searching for store constraint to build a complete match for a rule head, thus 
allowing for execution of the rule.

Below, we give a goal-based execution of the earlier communication buffer example.
{\small
\bda{c}
 get ~\atsign~ Get(x),Put(y) \simparrow x=y \\ \\
 
 \ba{lll}
     & & \chrstate{\{Get(x_1),Get(x_2),Put(1),Put(2)\}}{\{\}} \\
  \mbox{(D1 Activate)}  & \goaltrans & \chrstate{\{Get(x_1)\#1,Get(x_2),Put(1),Put(2)\}}{\{Get(x_1)\#1\}} \\
  \mbox{(D2 Drop)}      & \goaltrans & \chrstate{\{Get(x_2),Put(1),Put(2)\}}{\{Get(x_1)\#1\}} \\
  \mbox{(D3 Activate)}  & \goaltrans & \chrstate{\{Get(x_2)\#2,Put(1),Put(2)\}}{\{Get(x_1)\#1,Get(x_2)\#2\}} \\
  \mbox{(D4 Drop)}      & \goaltrans & \chrstate{\{Put(1),Put(2)\}}{\{Get(x_1)\#1,Get(x_2)\#2\}} \\
  \mbox{(D5 Activate)}  & \goaltrans & \chrstate{\{Put(1)\#3,Put(2)\}}{\{Get(x_1)\#1,Get(x_2)\#2,Put(1)\#3\}} \\
  \mbox{(D6 Fire } get) & \goaltrans & \chrstate{\{Put(2),x_1=1\}}{\{Get(x_2)\#2\}} \\
  \mbox{(D7 Activate}  & \goaltrans & \chrstate{\{Put(2)\#3,x_1=1\}}{\{Get(x_2)\#2,Put(2)\#3\}} \\
  \mbox{(D8 Fire } get) & \goaltrans & \chrstate{\{x_1=1,x_2=2\}}{\{\}} \\
  \mbox{(D9 Solve)}     & \goaltrans & \chrstate{\{x_2=2\}}{\{x_1=1\}} \\
  \mbox{(D10 Solve)}     & \goaltrans & \chrstate{\{\}}{\{x_1=1,x_2=2\}}
 \ea
\eda
}
We label the $x^{th}$ derivation step by a label $Dx$. 
Let's walk through each of the individual goal-based execution steps.
Initially, all constraints are kept in the set of goals. At this point,
all of the goals are inactive. Execution of goals proceeds in two stages:
(1) Activation and (2a) rule execution, or (2b) dropping of goals.
In the first stage, we activate a goal. In general, the order 
in which goals are activated is arbitrary.
For concreteness, we assume a left-to-right activation order.

Hence, we first activate $\Get(x_1)$ in derivation 
step (D1). Active goals carry a unique identifier, a distinct integer 
number. Besides assigning numbers to active goals, we also put them 
into the store. For instance, after activating $Get(x_1)$, we have
$Get(x_1)\#1$
in both the goals and the store.~\footnote{Numbered constraints also 
disambiguate multiple copies in the store but this is rather a side-effect.
The main purpose of numbering constraints is to indicate activation and retain
the link between active goal constraints and their stored copy.}

Active goals like $Get(x_1)\#1$ are executed by trying to build a complete
match for a rule head with matching partner constraints in the store.
Since there are no other constraints in the store, we cannot match
$Get(x_1)\#1$ with the $get$ rule. Therefore we drop $Get(x_1)\#1$ 
in step (D2). Dropping of a goal means the goal is removed from the set 
of goals but of course the (now inactive) goal is still present in 
the store.  Step (D3) and (D4) are similar but executed on  goal $Get(x_2)$.
Then, we activate $Get(x_2)$ and find that $Get(x_2)\#2$ cannot
build a complete match of the $get$ rule, thus it is dropped too.

Next, we activate $Put(1)$ (Step D5). Constraint $Put(1)\#3$ can match with
either $Get(x_1)\#1$ or $Get(x_2)\#2$ to form a complete instance of rule head 
of $get$. We pick $Get(x_1)\#1$ and fire the rule $get$, see step (D6). 
Step (D7) and (D8) perform similar execution steps on $Put(2)$ and the remaining
stored constraint $Get(x_2)\#2$. Finally, we add the equations
$x_1=1$ and $x_2=2$ into the store in steps (D9) and (D10). Exhaustive
application of this goal-based execution strategy then leads to a
state with no goals and a final store.

\fig{fig:conc-example}{Example of concurrent goal-based CHR derivation}{
{\tiny
\bda{cc}
\hspace{-15mm} &
\ba{l} 
 \myirule{
  \ba{c}
    \mbox{Short hands: } G = Get \sgap P = Put \\
    \chrstate{\{G(x_1),G(x_2),P(1),P(2)\}}{\{\}} \\ \\
    \ba{c}
    \mbox{(D1a Activate)} \sgap \goaltranssf{\{\} \backslash \{\}} 
                          \chrstate{\{G(x_1)\#1,G(x_2),P(1),P(2)\}}{\{G(x_1)\#1\}} \\
    \sgap \mid\mid \sgap \\
    \mbox{(D1b Activate)} \sgap \goaltranssf{\{\} \backslash \{\}} 
                          \chrstate{\{G(x_1),G(x_2)\#2,P(1),P(2)\}}{\{G(x_2)\#2\}}    
    \ea
  \ea
 }
 {
  \myirule{
   \ba{c}
    \ba{ll}
                                       & \chrstate{\{G(x_1),G(x_2),P(1),P(2)\}}{\{\}} \\
      \mbox{(D1a} \mid\mid \mbox{D1b)} \sgap \partranssf{\{\} \backslash \{\}} 
        & \chrstate{\{G(x_1)\#1,G(x_2)\#2,P(1),P(2)\}}{\{G(x_1)\#1,G(x_2)\#2\}}
    \ea \\ \\
    \ba{c}
    \mbox{(D2a Drop)} \sgap \goaltranssf{\{\} \backslash \{\}}
                            \chrstate{\{G(x_2)\#2,P(1),P(2)\}}{\{G(x_1)\#1,G(x_2)\#2\}} \\
    \sgap \mid\mid \sgap \\
    \mbox{(D2b Drop)} \sgap \goaltranssf{\{\} \backslash \{\}} 
                            \chrstate{\{G(x_1)\#1,P(1),P(2)\}}{\{G(x_1)\#1,G(x_2)\#2\}}
    \ea
   \ea
  }
  {
   \myirule{
    \ba{c}
     \ba{ll}
                & \chrstate{\{G(x_1)\#1,G(x_2)\#2,P(1),P(2)\}}{\{G(x_1)\#1,G(x_2)\#2\}} \\
      \mbox{(D2a} \mid\mid \mbox{D2b)} \sgap \partranssf{\{\} \backslash \{\}} 
        & \chrstate{\{P(1),P(2)\}}{\{G(x_1)\#1,G(x_2)\#2\}}
     \ea \\ \\
     \ba{c}
     \mbox{(D3a Activate)} \sgap \goaltranssf{\{\} \backslash \{\}}
        \chrstate{\{P(1)\#3,P(2)\}}{\{G(x_1)\#1,G(x_2)\#2,P(1)\#3\}} \\
     \sgap \mid\mid \sgap \\
     \mbox{(D3b Activate)} \sgap \goaltranssf{\{\} \backslash \{\}} 
        \chrstate{\{P(1),P(2)\#4\}}{\{G(x_1)\#1,G(x_2)\#2,P(2)\#4\}}
     \ea
    \ea
   }
   {
    \myirule{
     \ba{c}
      \ba{ll}
             & \chrstate{\{P(1),P(2)\}}{\{G(x_1)\#1,G(x_2)\#2\}} \\
       \mbox{(D3a} \mid\mid \mbox{D3b)} \sgap \partranssf{\{\} \backslash \{\}} 
          & \chrstate{\{P(1)\#3,P(2)\#4\}}{\{G(x_1)\#1,G(x_2)\#2,P(1)\#3,P(2)\#4\}}
      \ea \\ \\
      \ba{c}
       \mbox{(D4a Fire } get) \sgap \goaltranssf{\delta_1}
          \chrstate{\{x_1=1,P(2)\#4\}}{\{G(x_2)\#2,P(2)\#4\}} \\
       \sgap \mid\mid \sgap \\      
       \mbox{(D4b Fire } get) \sgap \goaltranssf{\delta_2} \chrstate{\{P(1)\#3,x_2=2\}}{\{G(x_1)\#1,P(1)\#3\}} \\
        \mbox{where} ~~ \delta_1 = \{\} \backslash \{G(x_1)\#1,P(1)\#3\} \sgap
       \delta_2 = \{\} \backslash \{G(x_2)\#2,P(1)\#4\} 
      \ea
     \ea
    }
    {
     \myirule{
      \ba{c}
       \ba{ll}
          & \chrstate{\{P(1)\#3,P(2)\#4\}}{\{G(x_1)\#1,G(x_2)\#2,P(1)\#3,P(2)\#4\}} \\
        \mbox{(D4a} \mid\mid \mbox{D4b)} \sgap \partranssf{\delta} & \chrstate{\{x_1=1,x_2=2\}}{\{\}}
       \ea \\ 
   \mbox{where}~~       \delta = \{\} \backslash \{G(x_1)\#1,P(1)\#3,G(x_2)\#2,P(1)\#4\}
\\
       \ba{c}
       \mbox{(D5a Solve)} \sgap \goaltranssf{\{\} \backslash \{\}} \chrstate{\{x_2=2\}}{\{x_1=1\}} 
       \sgap \mid\mid \sgap 
       \mbox{(D5b Solve)} \sgap \goaltranssf{\{\} \backslash \{\}} \chrstate{\{x_1=1\}}{\{x_2=2\}}
       \ea
      \ea
     }
     {
      \ba{ll}
          & \chrstate{\{x_1=1,x_2=2\}}{\{\}} \\
       \mbox{(D5a} \mid\mid \mbox{D5b)} \sgap \partranssf{\{\} \backslash \{\}} & \chrstate{\{\}}{\{x_1=1,x_2=2\}}
      \ea
     }
    }
   }
  }
 }
\ea
\eda
}
}

What we have described so far is essentially the execution scheme in which all 
major CHR implementations are based on. 
The semantics of these implementations assume a deterministic activation policy.
For example, goals are kept in a stack~\cite{DuckSBH04}
or priority queue~\cite{rp-chr}. This of course implies a strictly sequential
execution scheme.

To obtain a systematic, yet concurrent, CHR execution scheme we adapt
the goal-based CHR semantics as follows. Several active goal constraints
can simultaneously 
seek for partner constraints in the store to fire a rule instance. In the extreme 
case, all goal constraints could be activated at once. However, we generally assume 
that the number of active goals are bounded by $n$ where $n$ corresponds to 
the the number of actual threads available to the run-time system (for example, processor cores).

Figure \ref{fig:conc-example} shows a sample concurrent goal-based CHR derivation.
We assume two concurrent threads, referred to as $a$ and $b$, each thread executes the 
standard goal-based derivation steps.
The novelty is that each goal-based derivation step $\goaltranssf{\delta}$ now records
its effect on the store. The effect $\delta$ represents the sets of constraints in the 
store which were propagated or simplified.
Goal-based derivation steps can be executed concurrently if their effects
are not in conflict.

{\small
\bda{cc}
 \tlabel{\bf Goal-Concurrency} &
      \myirule{\chrstate{G_1}{H_{S1} \stcup H_{S2} \stcup S} \partranssf{\delta_1}
               \chrstate{G_1'}{H_{S2} \stcup S} \\
               \chrstate{G_2}{H_{S1} \stcup H_{S2} \stcup S} \partranssf{\delta_2} 
               \chrstate{G_2'}{H_{S1} \stcup S} \\
               \delta_1 = H_{P1} \backslash H_{S1} \sgap \delta_2 = H_{P2} \backslash H_{S2} \\
               H_{P1} \subseteq S \sgap H_{P2} \subseteq S \sgap \delta = H_{P1} \cup H_{P2} \backslash H_{S1} \cup H_{S2}}
              {\ba{ll}
                 & \chrstate{G_1 \uplus G_2 \uplus G}{H_{S1} \stcup H_{S2} \stcup S} \\
                \partranssf{\delta} 
                 & \chrstate{G_1' \uplus G_2' \uplus G}{S}
               \ea}
\eda }

The \tlabel{Goal-Concurrency} rule, abbreviated ($\parallel{\cal G}$), states that two
goal-derivations are not in conflict if their simplification effects are disjoint
and the propagated effects are present in the joint store.
We will provide more explanations later. Let's continue with our example.

Each thread activates one of the two $Get$ goals (Steps D1a and D1b). Since
both steps involve no rule application, side-effects are empty ($\{\} \backslash \{\}$).
Both steps are executed concurrently denoted by the concurrent 
derivation step (D1a$\mid\mid$D2a) $\partranssf{\{\} \backslash \{\}}$. 
Concurrent goal-based execution threads operate on a shared store and their effects will
be immediately made visible to other threads. This is important to guarantee exhaustive
rule firings. 

In the second step (D2a$\mid\mid$D2b),
both active goals are dropped because there is no complete match for any rule head yet.
Next, steps (D3a) and (D3b) activate the last two  goal constraints, $Put(1)$ and $Put(2)$. 
Each active constraint can match with either of the two $Get$ constraints in the store. 
We assume that active constraint $Put(1)\#3$ in step (D4a) matches with $Get(x_1)\#1$, while 
$Put(2)\#4$ in step (D4b) matches with $Get(x_2)\#2$, corresponding to the side-effects
$\delta_1$ and $\delta_2$. This guarantees that steps (D4a) and (D4b) operates on different 
(non-conflicting) parts of the store. Thus, we can execute them concurrently which yields 
step (D4a$\mid\mid$D4b). Their side-effects are combined as $\delta$. Finally, in 
step (D5a$\mid\mid$D5b) we concurrently solve the two remaining equations by adding
them into the store and we are done.

The correctness of our concurrent goal-based semantics is established by showing that
all concurrent derivations can be replicated by sequential goal-based executions. We
also prove that there is a correspondence between our goal-based CHR 
semantics with the abstract CHR semantics. This proof generalizes from \cite{greg:thesis}
which shows a correspondence between the refined CHR operational semantics and abstract
semantics. There are a number of subtle points we came across when developing the concurrent 
variant of the goal-based semantics. We will postpone a discussion of these issues, as well 
as a complete formalization of the concurrent goal-based semantics until Section
\ref{sec:concurrent-chr}. Next, we formally introduce the details of the abstract CHR semantics.


\section{Constraint Handling Rules} \label{sec:chr-example} \label{ssec:chr-sem}

\fig{fig:chrsemantics}{Abstract CHR semantics}{
\bda{l}
 \mbox{\bf Notations:} \\ 
 \ba{lllll}
  \uplus \bgap & \mbox{Multi-set union} \\
   \models \bgap & \mbox{Theoretic entailment} \\
  \phi         & \mbox{Substitution}   \\
   \overline{a}  & \mbox{Set/List of $a$'s} 
 \ea \\ \\
 \mbox{\bf CHR Syntax:} \\
 \ba{lllll}
  \mbox{Functions} \sgap & f ::= + \mid > \mid \&\& \mid ...  \\
  \mbox{Constants} \sgap & v ::= 1 \mid true \mid ... \\
  \mbox{Terms}           & t ::= x \mid f~\overline{t} \\
  \mbox{Predicates}      & p ::= Get \mid Put \mid ... \\
  \mbox{Equations}       & e ::= t = t \\
  \mbox{CHR constraints} & c ::= p(\overline{t}) \\
  \mbox{Constraints}     & b ::= e \mid c  \\
  \mbox{CHR Guards}      & t_g ::= t \\
  \mbox{CHR Heads}       & H ::= \overline{c} \\
  \mbox{CHR Body}        & B ::= \overline{b} \\
  \mbox{CHR Rule}        & R ::= r~\atsign ~H ~\backslash~ H \simparrow t_g \mid B \\
  \mbox{CHR Store}       & S ::= \overline{b} \\
  \mbox{CHR Program}     & {\cal P} ::= \overline{R}
 \ea \\ \\
 \mbox{\bf Abstract Semantics Rules:} \bgap\ruleform{\mathit{Store} \abstrans \mathit{Store}} \\ \\
 \ba{ccc}
 \tlabel{\bf Rewrite} & & 
  \myirule{(r~\atsign ~H_P \backslash H_S \simparrow t_g \mid B) \in {\cal P} \mbox{ such that } \\
          \exists \phi \sgap Eqs(S) \models \phi \wedge t_g \sgap \phi(H_P \uplus H_S) = H_P' \uplus H_S' }
          {H_P' \uplus H_S' \uplus S \abstrans H_P' \uplus \phi(B) \uplus S} \\ \\
 \tlabel{\bf Concurrency} & &  
  \myirule{S\uplus S_1 \abstransstar S \uplus S_2 \sgap 
           S\uplus S_3 \abstransstar S \uplus S_4}
          {S\uplus S_1 \uplus S_3 \abstransstar S \uplus S_2 \uplus S_4} \\ \\
   \tlabel{\bf Closure} & &
   \myirule{S \abstrans S'}{S \abstransstar S'} \bgap   
   \myirule{S \abstrans S' \sgap S' \abstransstar S''}
           {S \abstransstar S''}
 \ea \\ \\
 \ba{c}
    \bgap \bgap \bgap \bgap \mbox{where } Eqs(S) = \{e \mid e \in S, e \mbox{ is an equation}\}
 \ea 
\eda
}

Figure~\ref{fig:chrsemantics} reviews the essentials of the abstract CHR 
semantics~\cite{fruehwirth98:chr:art}. The general form of CHR rules contains propagated 
heads $H_P$ and simplified heads $H_S$ as well as a guard $t_g$
$$
r~\atsign ~H_P \backslash H_S \simparrow t_g \mid B
$$
In CHR terminology, a rule with simplified heads only ($H_P$ is empty) is referred to as
a {\em simplification} rule, a rule with propagated heads only ($H_S$ is empty) is referred to as
a {\em propagation} rule. The general form is referred to as a {\em simpagation} rule.

CHR rules manipulate a global constraint store which is a multi-set of constraints.
We execute CHRs by exhaustive rewriting of constraints in the store
with respect to the given rule system (a finite set of CHR rules), via the derivations
$\rightarrowtail$. To avoid ambiguities, we annotate derivations  of the abstract 
semantics with ${\cal A}$.

Rule \tlabel{Rewrite} describes application of a CHR rule \rlabel{r} at some instance $\phi$.
We simply (remove from the store)  the matching copies of $\phi(H_S)$ 
and propagate (keep in the store) the matching copies of $\phi(H_P)$.
But this only happens if the instantiated guard $\phi(t_g)$ is entailed by the equations 
present in the store $S$, written $Eqs(S) \models \phi(t_g)$.
In case of a propagation rule we need to avoid infinite re-propagation. We refer 
to~\cite{abdennadher:confluence,greg:thesis} for details. 
Rule \tlabel{Concurrency}, introduced in~\cite{union-find}, states that rules can be 
applied concurrently as long as they 
simplify on non-overlapping parts of the store.
\begin{definition} 
   [Non-overlapping Rule Application] \label{def:nonoverlap-rule}
   Two applications of the rule instances
   $r~\atsign~H_P \backslash H_S \simparrow t_g \mid B$ and 
   $r'~\atsign~H_P' \backslash H_S' \simparrow t_g' \mid B'$ in store $S$ are said to 
   be non-overlapping if and only if they simplify unique parts of $S$ 
   (i.e. $H_S,H_S' \subseteq S$ and $H_S \cap H_S' = \emptyset$).
\end{definition}
The two last \tlabel{Closure} rules simply specify the transitive application of CHR rules.

\ignore{
We conclude this section by some elementary definition and results.

It is a consequence of the {\em monotonicity} of CHR (Theorem \ref{theo:monotonicity},
proven in \cite{Abdennadher99confluenceand}) which states that rule applications will
remain valid in a larger context (store). In other words, presence of additional constraints
in the store will not prevent the firing of rules. 

Note that we consider a minor limitation on the equations we use: We treat equations as
variable assignment. Hence equations used in CHR rule body are only of the form
$x = v$, where $v$ is strictly a value or function application.
}

\section{Concurrent Goal-Based CHR Operational Semantics} \label{sec:concurrent-chr} \label{ssec:goal-execution}


\fig{fig:chr-goal-syntax}{CHR Goal-based Syntax}{
\bda{l}
 \mbox{\bf Notations:} \\ 
 \ba{lllll}
  \uplus \bgap & \mbox{Multi-set union} \\
  \cup   \bgap & \mbox{Set union} \\ 
  \models \bgap & \mbox{Theoretic entailment} \\
  \phi         & \mbox{Substitution}   \\ 
  \overline{a}  & \mbox{Set/List of $a$'s} 
 \ea \\ \\
 \mbox{\bf CHR Syntax:} \\
 \ba{lllll}
  \mbox{Functions} \sgap & f ::= + \mid > \mid \&\& \mid ...  \\
  \mbox{Constants} \sgap & v ::= 1 \mid true \mid ... \\
  \mbox{Terms}           & t ::= x \mid f~\overline{t} \\
  \mbox{Predicates}      & p ::= Get \mid Put \mid ... \\
  \mbox{Equations}       & e ::= t = t \\
  \mbox{CHR Constraints} & c ::= p(\overline{t}) \\
  \mbox{Constraints}     & b ::= e \mid c  \\
  \mbox{CHR Guards}      & t_g ::= t \\
  \mbox{CHR Heads}       & H ::= \overline{c} \\
  \mbox{CHR Body}        & B ::= \overline{b} \\
  \mbox{CHR Rule}        & R ::= r~\atsign ~H ~\backslash~ H \simparrow t_g \mid B \\
  \mbox{CHR Program}     & {\cal P} ::= \overline{R} \\
  \mbox{Num Constraint} & nc ::= c\#i         \\ \mbox{Goal Constraint} & g ::= c \mid e \mid nc \\
  \mbox{Stored Constraint} & sc ::= nc \mid e \\ \mbox{CHR Num Store} & Sn ::= \overline{sc} \\
  \mbox{CHR Goals} & G ::= \overline{g}       \\ \mbox{CHR State} & \sigma ::= \langle G,Sn \rangle \\
  \mbox{Side Effects} & \delta ::= Sn ~\backslash~ Sn
  \ea
\eda
}

We present the formal details of the concurrent goal-based CHR semantics. Figure 
\ref{fig:chr-goal-syntax} describes the necessary syntactic extensions. Because
constraints in the store now have unique identifiers, we treat the store as a
set (as opposed to a multiset) and use set union $\stcup$. Goals are still treated as multi-sets
because they can contain multiple copies of (un-numbered) CHR constraints.
The actual semantics is given in two parts. Figure~\ref{fig:goal-semantics} 
describes the single-step execution part whereas
Figure \ref{fig:conc-goal-semantics} introduces the concurrent execution part.
The first part is a generalization of an earlier goal-based description~\cite{greg:thesis}
whereas the second (concurrent) part is novel.


\fig{fig:goal-semantics}{Goal-Based CHR Semantics (Single-Step Execution)}{
\bda{c} 
 \ba{c}
       \ruleform{\chrstate{\mathit{Goal}}{\mathit{Store}} 
                 \goaltranssf{\delta} 
                 \chrstate{\mathit{Goal}}{\mathit{Store}}}
 \ea
  \\ \\
   \ba{ccc}
      \tlabel{\bf Solve} & & 
      \myirule{W = WakeUp(e,Sn)}
      {\chrstate{\{e\} \uplus G}{Sn} \goaltranssf{W \backslash \{\}} \chrstate{W \uplus G}{\{e\} \stcup Sn}} \\ \\
      \tlabel{\bf Activate} & &
      \myirule{i \mbox{ is a fresh identifier}}
              {\chrstate{\{c\} \uplus G}{Sn} \goaltranssf{\{\} \backslash \{\}} \chrstate{\{c\#i\} \uplus G}{\{c\#i\} \stcup Sn}} \\ \\
      \tlabel{\bf Simplify} & &
      \myirule{ (r~\atsign~H_P' \backslash H_S' \simparrow t_g \mid B') \in {\cal P} \mbox{ such that} \\
                \exists \phi \sgap Eqs(Sn) \models \phi \wedge t_g \sgap \phi(H_P') = DropIds(H_P) \\
                        \phi(H_S') = \phi(\{c\} \uplus DropIds(H_S)) \\
                \delta = H_P \backslash \{c\#j\} \stcup H_S }
              {\ba{ll}
                & \chrstate{\{c\#j\} \uplus G}{\{c\#j\} \stcup H_P \stcup H_S \stcup Sn} \\
               \goaltranssf{\delta} & \chrstate{\phi(B') \uplus G}{H_S \stcup Sn}
               \ea } \\ \\
      \tlabel{\bf Propagate} & &
      \myirule{ (r~\atsign~H_P' \backslash H_S' \simparrow t_g \mid B') \in {\cal P} \mbox{ such that} \\
                \exists \phi \sgap Eqs(Sn) \models \phi \wedge t_g \sgap \phi(H_S') = DropIds(H_S) \\
                        \phi(H_P') = \phi(\{c\} \uplus DropIds(H_P)) \\
                \delta = \{c\#j\} \stcup H_P \backslash H_S }
              {\ba{ll}
                & \chrstate{\{c\#j\} \uplus G}{\{c\#j\} \stcup H_P \stcup H_S \stcup Sn} \\
               \goaltranssf{\delta} & \chrstate{\phi(B') \uplus \{c\#j\} \uplus G} 
                                               {\{c\#j\} \stcup H_P \stcup Sn}
               \ea } \\ \\
      \tlabel{\bf Drop} & &
      \myirule{\tlabel{Simplify} \mbox{ and } \tlabel{Propagate} \mbox{ does not apply on } c\#j \mbox{ in } Sn}
              {\chrstate{\{c\#j\} \uplus G}{Sn} \goaltranssf{\{\} \backslash \{\}} 
               \chrstate{G}{Sn}}
   \ea
 \\ \\   
   \ba{clll}
   \mbox{where} 
         & Eqs(S)       & = & \{e \mid e \in S, e \mbox{ is an equation}\} \\       
         & DropIds(Sn)  & = & \{ c \mid c\#i \in Sn \} \uplus \{ e \mid e \in Sn, e \mbox{ is an equation}\} \\
         & WakeUp(e,Sn) & = & \{ c\#i \mid c\#i \in Sn \wedge \phi \mbox{ m.g.u. of } Eqs(Sn) \wedge \\
         &              &   & \theta \mbox{ m.g.u. of } Eqs(Sn \cup \{e\}) \wedge \phi(c) \neq \theta(c) \} 
   \ea
\eda
}

We first discuss the single-step derivation steps in Figure~\ref{fig:goal-semantics}. A derivation step
$\sigma \goaltranssf{\delta} \sigma'$ maps the CHR state $\sigma$ to $\sigma'$ with some side-effect $\delta$. 
$\delta$ represents the constraints that where propagated or simplified during rule application. Hence
derivation steps that do not involve rule application (\tlabel{Activate} and \tlabel{Drop})
contain no side-effects (i.e. $\{\} \backslash \{\}$). We will omit side-effects $\delta$ as and when it is not
relevant to our discussions. We ignore the \tlabel{Solve} step for the moment.
In \tlabel{Activate}, we activate a goal CHR constraint by assigning it a fresh unique identifier and adding it 
to the store. Rewrite rules are executed in steps \tlabel{Simplify} and \tlabel{Propagate}. We distinguish whether 
the rewrite rule is executed on a simplified or propagated  active (goal) constraint $c\#i$. For both cases, 
we seek for the missing partner constraints in the store for some matching substitution $\phi$. The auxiliary 
function $DropIds$ ignores the unique identifiers of numbered constraints. They do not matter when finding a 
rule head match. The guard $t_g$ must be entailed by the primitive (here equational) store constraints under the 
substitution $\phi$.

In case of a simplified goal, step \tlabel{Simplify}, we apply the rule instance of $r$ by deleting all 
simplified matching constraints $H_S$ and adding the rule body instance $\phi(B)$ into the goals. Since 
$c\#i$ is simplified, we drop $c\#i$ from the goals as it does not exist in the store any more. In case of 
a propagated goal, step \tlabel{Propagate}, $c\#i$ remains in the goal set as well in the store and thus 
can possibly fire further rules instances. For both \tlabel{Simplify} and \tlabel{Propagate} derivation
step, say $\sigma \goaltranssf{H_P \backslash H_S} \sigma'$, we record as side-effect the numbered constraints 
in the store that were propagated ($H_P$) or simplified ($H_S$) during the derivation step. We will elaborate 
on the purpose of side-effects when we introduce the concurrent part of the semantics.

In step \tlabel{Drop}, we remove an active constraint from the set of goals, 
if the constraint failed to trigger any CHR rule.

Rule \tlabel{Solve} moves an equation goal $e$ into the store and {\em wakes up} (reactivates)
any numbered constraint in the store which can possibly trigger further CHR rules due to the presence of $e$. 
Here is a simple example to show why reactivation is necessary.
{\small
\bda{l}
  r1~\atsign~A(x),B(x)\simparrow C(x) \\ \\
  \ba{lll}
         & & \chrstate{\{a=2\}}{\{A(a)\#1,B(2)\#2\}} \\
    \mbox{(Solve)} \sgap & \goaltranssf{\{A(2)\#1\} \backslash \{\}} 
         & \chrstate{\{A(2)\#1\}}{\{A(2)\#1,B(2)\#2,a=2\}} \\
    \mbox{(Simp } r1) \sgap & \goaltranssf{\{\} \backslash \{A(2)\#1,B(2)\#2\}}
         & \chrstate{\{C(2)\}}{\{a=2\}} \\
    ...
  \ea 
 \eda }
For clarity, we normalize all constraints in the store once 
an equation is added. Prior to addition of $a=2$, $A(a)\#1,B(2)\#2$ cannot fire rule $r1$. 
After adding $a=2$ however, we can normalize $A(a)\#1$ to $A(2)\#2$, which can now fire $r1$ 
with $B(2)\#2$. To guarantee exhaustive rule firings, we reactivate $A(2)\#2$ by adding it back 
to the set of goals. $WakeUp(e,Sn)$ represents a conservative approximation
of the to be reactivated constraints~\cite{greg:thesis}. Note that we treat reactivated constraints
as propagated constraints in the side-effects.

\fig{fig:conc-goal-semantics}{Goal-Based CHR Semantics (Concurrent Part)}{
 \bda{c}
 \ba{c}
       \ruleform{\chrstate{\mathit{Goal}}{\mathit{Store}} \partranssf{\delta} \chrstate{\mathit{Goal}}{\mathit{Store}}} 
 \ea 
  \\ \\
   \ba{cc}
      \tlabel{\bf Lift} &
      \myirule{\chrstate{G}{Sn} \goaltranssf{\delta} \chrstate{G'}{Sn'}}
              {\chrstate{G}{Sn} \partranssf{\delta} \chrstate{G'}{Sn'}} \\ \\
      \tlabel{\bf Goal Concurrency} & 
      \myirule{\chrstate{G_1}{H_{S1} \stcup H_{S2} \stcup S} \partranssf{\delta_1}
               \chrstate{G_1'}{H_{S2} \stcup S} \\
               \chrstate{G_2}{H_{S1} \stcup H_{S2} \stcup S} \partranssf{\delta_2} 
               \chrstate{G_2'}{H_{S1} \stcup S} \\
               \delta_1 = H_{P1} \backslash H_{S1} \sgap \delta_2 = H_{P2} \backslash H_{S2} \\
               H_{P1} \subseteq S \sgap H_{P2} \subseteq S \sgap \delta = H_{P1} \cup H_{P2} \backslash H_{S1} \cup H_{S2}}
              {\ba{ll}
                 & \chrstate{G_1 \uplus G_2 \uplus G}{H_{S1} \stcup H_{S2} \stcup S} \\
                \partranssf{\delta} 
                 & \chrstate{G_1' \uplus G_2' \uplus G}{S}
               \ea} \\ \\   
   \tlabel{\bf Closure} &
   \myirule{\sigma \partranssf{\delta} \sigma'}{\sigma \partransstar \sigma'} \bgap   
   \myirule{\sigma \partranssf{\delta} \sigma' \sgap \sigma' \partransstar \sigma''}
           {\sigma \partransstar \sigma''}
   \ea
 \eda
}

Figure \ref{fig:conc-goal-semantics} presents the concurrent part of the goal-based operational 
semantics. In the \tlabel{Lift} step, we turn a sequential goal-based derivation into a 
concurrent derivation. Note that side-effects are retained. Step \tlabel{Goal Concurrency} joins 
together two concurrent derivations operating on a shared store, if
their rewriting side-effects $\delta_1$ and $\delta_2$ are 
non-overlapping as defined below.

\begin{definition} [Non-overlapping Rewriting Side-Effects] \label{def:non-overlap-side-effects}
   Two rewriting side-effects $\delta_1 = H_{P1} \backslash H_{S1}$ and 
   $\delta_2 = H_{P2} \backslash H_{S2}$ are said to be non-overlapping,
   if and only if $H_{S1} \cap (H_{P2} \cup H_{S2}) = \{\}$ and 
   $H_{S2} \cap (H_{P1} \cup H_{S1}) = \{\}$ 
\end{definition}

Concurrent derivations with non-overlapping side-effects essentially simplify distinct constraints
in the store, as well as propagate constraints which are not simplified by one another. The
\tlabel{Goal Concurrency} step expresses non-overlapping side-effects by structurally enforcing that
simplified constraints $H_{S1}$ and $H_{S2}$ match distinct parts of the store, while propagated
constraints $H_{P1}$ and $H_{P2}$ are found in the shared part of the store $S$ not modified by 
both concurrent derivations. In the resulting concurrent derivation, the side-effects $\delta_1$ 
and $\delta_2$ are composed by the union of the propagate and simplify components respectively, 
forming $\delta$.

An immediate consequence is that we can execute $k$ derivations concurrently by stacking them 
together as long as all side-effects are mutually non-overlapping.
The following lemma summarizes this observation.

\begin{lemma}[$k$-Concurrency] \label{lem:k-concurrency}
For any finite $k$ of mutually non-overlapping concurrent derivations,
{\small
\bda{c}
  \myirule
  {\ba{c}
   \chrstate{G_1}{H_{S1} \stcup .. \stcup H_{Si} \stcup .. \stcup H_{Sk} \stcup S}
   \partranssf{H_{P1} \backslash H_{S1}}
   \chrstate{G_1'}{\{\} \stcup .. \stcup H_{Si} \stcup .. \stcup H_{Sk} \stcup S} \\
   ... \\
   \chrstate{G_i}{H_{S1} \stcup .. \stcup H_{Si} \stcup .. \stcup H_{Sk} \stcup S}
   \partranssf{H_{Pi} \backslash H_{Si}}
   \chrstate{G_i'}{H_{S1} \stcup .. \stcup \{\} \stcup .. \stcup H_{Sk} \stcup S} \\
   ... \\
   \chrstate{G_k}{H_{S1} \stcup .. \stcup H_{Si} \stcup .. \stcup H_{Sk} \stcup S}   
   \partranssf{H_{Pk} \backslash H_{Sk}}
   \chrstate{G_k'}{H_{S1} \stcup .. \stcup H_{Si} \stcup .. \stcup \{\} \stcup S} \\
   H_{P1} \subseteq S .. H_{Pi} \subseteq S .. H_{Pk} \subseteq S \\
   \delta = H_{P1} \cup .. \cup H_{Pi} \cup .. \cup H_{Pk} \backslash
            H_{S1} \cup .. \cup H_{Si} \cup .. \cup H_{Sk}
   \ea}
  {\ba{ll}
      & \chrstate{G_1 \uplus .. \uplus G_i \uplus .. \uplus G_k \uplus G}
                 {H_{S1} \stcup .. \stcup H_{Si} \stcup .. \stcup H_{Sk} \stcup S} \\
    \partranssf{\delta}
      & \chrstate{G_1' \uplus .. \uplus G_i' \uplus .. \uplus G_k' \uplus G}
                 {S}
   \ea}
\eda
}
we can decompose this into $k-1$ applications of the (pair-wise) \tlabel{Goal Concurrency} derivation step.
\end{lemma}

The \tlabel{Closure} step  defines transitive application of the concurrent goal-based 
derivation. Because side-effect labels are only necessary for the \tlabel{Goal Concurrency} step,
we drop the side-effects in transitive derivations.

Any concurrent goal-based derivation can be reproduced in the abstract CHR semantics.
This correspondence result is important to make use of the concurrent goal-based semantics
as a more systematic execution scheme for CHR.
We will formally verify this as well as other results in the up-coming Section~\ref{ssec:corr}.
First, we give an in-depth discussion of the more subtle aspects 
of the concurrent goal-based semantics.

\subsection{Discussion}

Most of the issues we encounter are related to the problem of exhaustive rule firings.
For brevity, we omit side-effects in derivation steps in the following examples as
they do not matter.

\paragraph{\bf Goal Storage, Shared Store and Single-Step Execution:}

\comment{
Our \tlabel{\bf Goal Concurrency} demands that concurrent executions operate on a shared store
and only make a single-step in the goal-based semantics. Furthermore, goals are immediately 
stored after activation. All three conditions are essential to ensure that we exhaustively 
fire rules, as the following examples demonstrate. Note for examples in this section, we 
omit side-effects as they are not necessary for their discussion.
}

Each of these issues affect (exhaustive) rule firings.
We first consider goal storage.
Suppose we would only store goals after execution (rule head matching).
That is, we do not add the goals into the store during \tlabel{Activate} step, but
only during the \tlabel{Drop} step.

{\small
\bda{c}
 \ba{ccc}
   \tlabel{Activate'} & & 
   \myirule{i \mbox{ is a fresh identifier}}
           {\chrstate{\{c\}\uplus G}{Sn} \goaltrans \chrstate{\{c\#i\}\uplus G}{Sn}} \\
   \tlabel{Drop'} & &
   \myirule{\tlabel{Simplify} \mbox{ and } \tlabel{Propagate} 
            \mbox{ does not apply on } c\#i \mbox{ in } Sn}
           {\chrstate{\{c\#i\} \uplus G}{Sn} \goaltrans \chrstate{G}{\{c\#i\} \stcup Sn}}
 \ea 
\eda
}

Then, for the CHR program
\bda{c}
 r1 ~\atsign~ A(x),B(y) \simparrow C(x,y) 
\eda
we obtain the following derivation
{\small
\bda{c}
   \myirule
   {\ba{c}
    \chrstate{\{A(1),B(2)\}}{\{\}} \\ \\
    \tlabel{Activate'} \sgap \chrstate{\{A(1)\}}{\{\}} \partrans \chrstate{\{A(1)\#1\}}{\{\}} \\
    \mid\mid \\
    \tlabel{Activate'} \sgap \chrstate{\{B(2)\}}{\{\}} \partrans \chrstate{\{B(2)\#2\}}{\{\}}
    \ea}
   {\chrstate{\{A(1),B(2)\}}{\{\}} \partrans \chrstate{\{A(1)\#1,B(2)\#2\}}{\{\}}} \\ \\
   \myirule
   {\ba{c}
    \tlabel{Drop'} \sgap \chrstate{\{A(1)\#1\}}{\{\}} \partrans \chrstate{\{\}}{\{A(1)\#1\}} \\
    \mid\mid \\
    \tlabel{Drop'} \sgap \chrstate{\{B(2)\#2\}}{\{\}} \partrans \chrstate{\{\}}{\{B(2)\#2\}} 
    \ea}
   {\chrstate{\{A(1)\#1,B(2)\#2\}}{\{\}} \partrans 
    \chrstate{\{\}}{\{A(1)\#1,B(2)\#2\}}}
  
\eda
}

Initially
both goals $A(1)$ and $B(2)$ are concurrently activated. Since \tlabel{Activate'} does not store goals
immediately, both active goals are not visible to each other in the store.
Hence, we wrongfully apply the \tlabel{Drop'} step for
both goals.
However, there is clearly  a complete rule head match $A(1)\#1,B(2)\#2$.

Next, we investigate the shared store issue.
Suppose we allow for concurrent executions on (non-shared) split stores.
Then, the following derivation is possible.
{\small
\bda{c}
   r1~\atsign~A,B \simparrow C \bgap r2~\atsign~D,E \simparrow F \\ \\
   \myirule{\mbox{(Drop)}~~ \chrstate{\{A\#3\}}{\{A\#3,E\#2\}} \partrans
            \chrstate{\{\}}{\{A\#3,E\#2\}}  \\
            \mbox{(Drop)}~~ \chrstate{\{D\#4\}}{\{B\#1,D\#4\}} \partrans
            \chrstate{\{\}}{\{B\#1,D\#4\}}}
           {\chrstate{\{A\#3,D\#4\}}{\{A\#3,B\#1,D\#4,E\#2\}} \partrans
            \chrstate{\{\}}{\{A\#3,B\#1,D\#4,E\#2\}}}
\eda
}
The resulting store is a final store, there are 
no more goals left. However, if we consider the entire store $\{A\#3,E\#2,B\#1,D\#4\}$, it is clearly
that goal $A\#3$ can execute rule \rlabel{r1} and goal $D\#4$ can execute rule \rlabel{r2}.
We conclude that splitting of the store leads to "stuck" states. We fail
to exhaustively fire CHR rules.

For similar reasons, we demand that when joining concurrent executions, each individual
execution can only make a single-step. Otherwise, we encounter again a stuck state.
\bda{c}
   r1~\atsign~A,B \simparrow C \\ \\
   \myirule{(P1) \sgap \chrstate{\{A\}}{\{ \}} \partrans 
            \chrstate{\{A\#2\}}{\{A\#2\}} \partrans 
            \chrstate{\{\}}{\{A\#2\}} \\
            (P2) \sgap \chrstate{\{B\}}{\{ \}} \partrans 
            \chrstate{\{B\#3\}}{\{ B\#3\}} \partrans
            \chrstate{\{\}}{\{B\#3\}} }
           {\chrstate{\{A,B\}}{\{ \}} \partransstar 
            \chrstate{\{\}}{\{A\#2,B\#3\}} }
\eda
The sequence of derivation steps (P1) first activates $A$ which is then dropped.
Similarly, (P2) activates $B$ which is then dropped as well which then leads to 
the stuck state $\chrstate{\{\}}{\{A\#2,B\#3\}}$. We clearly missed to fire rule 
\rlabel{r1}. This shows that single-step concurrent execution are essential
to guarantee that newly added constraints are visible to all concurrent active 
goals, hence we have exhaustive rule firings in the goal-based semantics.

The underlying reason for non-exhaustive firing of rules is that the goal-based semantics
is not monotonic in its store argument. However, execution is monotonic 
in the goal argument which leads us to the next issue.

\paragraph{\bf Lazy Matching and Asynchronous Goal Execution:}

When executing goals, we lazily compute only matches that contain the specific 
goal and immediately apply such matches without concerning any further matches.
For instance consider the following CHR program and goal-based derivation:
{\small
\bda{c} 
  r0~\atsign~A(x),B(y) \simparrow D(x,y) \\ \\
  \ba{ll}
     & \chrstate
        {\{A(1)\#4\} \uplus \{A(2),A(3)\}}
        {\{B(2)\#1,B(3)\#2,B(4)\#3,A(1)\#4\}} \\
   \goaltrans &
   \chrstate
    {\{D(1,2)\} \uplus \{A(2),A(3)\}}
    {\{B(3)\#2,B(4)\#3\}}
  \ea
\eda
}
We have applied the rule instance $A(1)\#4,B(2)\#1$ independently
of the existence of the other goals (i.e. $\{A(2),A(3)\}$). In the literature, such a matching 
scheme is known as a {\em lazy} matching scheme, and often implemented by variants of 
the LEAPS algorithm \cite{leaps90}. 

Lazy matching in the goal-based semantics is possible only because the goal-based semantics 
is monotonic with respect to the set of goals. The following illustrates this monotonicity 
property of goals:
{\small
\bda{c}
  \myirule
  {\ba{ll}
     & \chrstate
        {\{A(1)\#4\}}
        {\{B(2)\#1,B(3)\#2,B(4)\#3,A(1)\#4\}} \\
   \goaltrans &
   \chrstate
    {\{D(1,2)\}}
    {\{B(3)\#2,B(4)\#3\}}
   \ea}
  {  
   \ba{ll}
     & \chrstate
        {\{A(1)\#4\} \uplus \{A(2),A(3)\}}
        {\{B(2)\#1,B(3)\#2,B(4)\#3,A(1)\#4\}} \\
   \goaltrans &
   \chrstate
    {\{D(1,2)\} \uplus \{A(2),A(3)\}}
    {\{B(3)\#2,B(4)\#3\}}
  \ea} \\ \\
  \myirule{\chrstate{G}{Sn} \goaltrans \chrstate{G'}{Sn'}}
          {\chrstate{G \uplus G''}{Sn} \goaltrans \chrstate{G' \uplus G''}{Sn'}}
\eda
}
The above property essentially states that we can execute goals $G$ without prior
knowledge of goals $G''$. Because of monotonicity, we are guaranteed that future 
executions of $G''$ will not invalidate them. 

Monotonicity of the goals also allows us to
execute goals asynchronously. For instance, consider the following:
{\small
\bda{c}
  r1~\atsign~A(x),B(y) \simparrow C(x,y) \\ \\
  \myirule
  {\ba{ll}
     & \chrstate{\{A(1)\#1\}}{\{A(1)\#1,B(2)\#2\} \stcup \{A(3)\#3,B(4)\#4\}} \\
    \partranssf{\delta_1}
     & \chrstate{\{C(1,2)\}}{\{\} \stcup \{A(3)\#3,B(4)\#4\}} \\
     & \chrstate{\{A(3)\#3\}}{\{A(1)\#1,B(2)\#2\} \stcup \{A(3)\#3,B(4)\#4\}} \\
    \partranssf{\delta_2}
     & \chrstate{\{C(3,4)\}}{\{A(1)\#1,B(2)\#2\} \stcup \{\}} 
   \ea \\
   \delta_1 = \{\} \backslash \{A(1)\#1,B(2)\#2\} \sgap 
   \delta_2 = \{\} \backslash \{A(3)\#3,B(4)\#4\} \\
   \delta = \{\} \backslash \{A(1)\#1,B(2)\#2,A(3)\#3,B(4)\#4\}}
  {\ba{ll}
     & \chrstate{\{A(1)\#1\} \uplus \{A(3)\#3\}}{\{A(1)\#1,B(2)\#2\} \stcup \{A(3)\#3,B(4)\#4\}} \\
    \partranssf{\delta}
     & \chrstate{\{C(1,2)\} \uplus \{C(3,4)\}}{\{\} \stcup \{\}}
   \ea}
\eda
}
The above describes the concurrent execution of goals $A(1)\#1$ 
and $A(3)\#3$. Notice that in the derivations of the premise, we can ignore all goals 
which are not relevant to the derivation. For instance, execution of $A(1)\#1$ does 
not need goal $A(3)\#3$ to be visible, hence the goals effectively executes 
asynchronously. Goals do however, implicitly "synchronize" via the shared store.
Namely, concurrent derivations must be chosen such that rewrite side-effects
involve distinct parts of the store. 

\subsection{Correspondence Results} \label{ssec:corr}

We formally verify that the concurrent goal-based semantics is in exact correspondence
to the abstract CHR semantics when it comes to termination and exhaustive rule firings.
Detailed proofs are given in the appendix. In the main text, we provide key lemmas 
and proof sketches.
We first introduce some elementary definitions before stating the formal results.

\paragraph{\bf Definitions:}

The first two definitions concern the abstract CHR semantics.
A store is final if no further rules are applicable.

\begin{definition} [Final Store] \label{def:final-store}
    A store $S$ is known as a final store, denoted $Final_{\cal A}(S)$
   if and only if no more CHR rules applies on it (i.e. $\neg \exists S'$ such that
   $S \rightarrowtail_{\cal A} S'$).
\end{definition}

A CHR program terminates if all derivations lead to a final store in 
a finite number of states.

\begin{definition} [Terminating CHR Programs] \label{def:terminate}
  A CHR program $\cal P$ is said to be terminating, if and only if for any CHR store $S$,
  all derivations starting from $S$ are finite.
\end{definition}

Next, we introduce some definitions in terms of the goal-based semantics. In an initial state, all 
constraints are goals and the store is empty. Final states are states which no longer have any 
goals. We will prove the exhaustiveness of the goal-based semantics by proving a correspondence
between final stores in the abstract semantics and final states of the goal-based semantics

\begin{definition} [Initial and Final CHR States] \label{def:init-fin-state}
   An initial CHR state is a CHR state of the form $\chrstate{G}{\{\}}$ where $G$ 
   contains no numbered constraints ($c\#n$), while a final CHR state is of the form 
   $\chrstate{\{\}}{Sn}$
\end{definition}

A state is reachable if there exists a (sequential) goal-based sequence of derivations
to this state. We write $\goaltransstar$ to denote the transitive closure of $\goaltrans$.

\begin{definition} [Sequentially Reachable CHR states] \label{def:seq-reach}
   For any CHR program ${\cal P}$, a CHR state $\chrstate{G'}{Sn'}$ is said to be sequentially 
   reachable by ${\cal P}$ if and only if there exists some initial CHR state $\chrstate{G}{\{\}}$ 
   such that $\chrstate{G}{\{\}} \goaltransstar \chrstate{G'}{Sn'}$.
\end{definition}

\subsubsection{Correspondence of Derivations}

We build a correspondence between the abstract semantics and the concurrent goal-based
semantics. We begin with Theorem \ref{theo:corr-goal-trans}, which states the correspondence
of the (sequential) goal-based semantics.

\begin{theorem} [Correspondence of Sequential Derivations] \label{theo:corr-goal-trans}
   For any reachable CHR state $\chrstate{G}{Sn}$, CHR state $\chrstate{G'}{Sn'}$ and 
   CHR program ${\cal P}$, 
   \bda{ll} 
    \mbox{if }   & \chrstate{G}{Sn} \goaltransstar \chrstate{G'}{Sn'} \\
    \mbox{then } & (NoIds(G) \uplus DropIds(Sn)) = (NoIds(G') \uplus DropIds(Sn')) \sgap \vee \\
                 & (NoIds(G) \uplus DropIds(Sn)) \abstransstar (NoIds(G') \uplus DropIds(Sn'))
   \eda
   where $NoIds = \{c \mid c \in G, c \mbox{ is a CHR constraint}\} 
          \uplus \{e \mid e \in G, e \mbox{ is an equation}\}$
\end{theorem}

The above result guarantees that any sequence of sequential goal-based derivations starting from a
reachable CHR state either yields equivalent CHR abstract stores (due to goal-based behavior
not captured by the abstract semantics, namely \tlabel{Solve} \tlabel{Activate}, 
\tlabel{Drop}) or corresponds to a derivation in the abstract semantics (due to rule 
application). A goal-based semantics state $\chrstate{G}{Sn}$ is related to an abstract 
semantics store by removing all numbered constraints in $G$ and unioning it with constraints
in $Sn$ without their identifiers. The theorem and its proof is a generalization 
of an earlier result given in \cite{greg:thesis}.

We formalize the observation that the goal context can
be extended without interfering with previous goal executions. 

\begin{lemma}[Monotonicity of Goals in Goal-based Semantics] \label{lem:mono-goals}
   For any goals $G$,$G'$ and $G''$ and CHR store $Sn$ and $Sn'$,
   If $\chrstate{G}{Sn} \goaltransstar \chrstate{G'}{Sn'}$
   then $\chrstate{G \uplus G''}{Sn} \goaltransstar \chrstate{G' \uplus G''}{Sn'}$.
\end{lemma}

Next, we state that
given any goal-based derivation with side-effects
$\delta$, we can safely ignore any constraints (represented by $S_2$) in the store
which is not part of $\delta$. 

\begin{lemma}[Isolation of Goal-based Derivations] \label{lem:isolation}
  If
     ${\chrstate{G}{H_P \stcup H_S \stcup S_1 \stcup S_2} 
      \goaltranssf{H_P \backslash H_S} 
      \chrstate{G'}{H_P \stcup S_1' \stcup S_2}
     }$
\\
  then
     ${\chrstate{G}{H_P \stcup H_S \stcup S_1} 
      \goaltranssf{H_P \backslash H_S} 
      \chrstate{G'}{H_P \stcup S_1'}}$
\end{lemma}

Lemma \ref{lem:isolation} can be straight-forwardly extended to multiple derivation
steps. This is stated in Lemma \ref{lem:isolationstar}.

\begin{lemma}[Isolation of Transitive Goal-based Derivations] \label{lem:isolationstar}
 If
     ${\chrstate{G}{H_P \stcup H_S \stcup S_1 \stcup S_2} 
      \goaltransstar
      \chrstate{G'}{H_P \stcup S_1' \stcup S_2} \\
      \mbox{with side-effects } \delta = H_P \backslash H_S
     }$
  then
     ${\chrstate{G}{H_P \stcup H_S \stcup S_1} 
      \goaltransstar 
      \chrstate{G'}{H_P \stcup S_1'}}$
\end{lemma}

The next states that any concurrent derivation starting from a reachable CHR state
can be replicated by a sequence of sequential goal-based derivations.
Lemma \ref{lem:corr-par-trans} is the first step to prove the correspondence of
concurrent goal-based derivations.

\begin{lemma} [Sequential Reachability of Concurrent Derivation Steps] \label{lem:corr-par-trans}
   For any sequentially reachable CHR state $\sigma$, CHR state $\sigma'$ and rewriting
   side-effects $\delta$ if $\sigma \partranssf{\delta} \sigma'$ then $\sigma'$ is 
   sequentially reachable, $\sigma \goaltransstar \sigma'$ with side-effects $\delta$.
\end{lemma}
\begin{proof}(Sketch)
Via Lemma \ref{lem:k-concurrency}, we can always reduce $k$ mutually non-overlapping concurrent 
derivations into several applications of the \tlabel{Goal Concurrency} step. Hence we
can prove Lemma \ref{lem:corr-par-trans} by structural induction over the
concurrent goal-based derivation steps \tlabel{Lift} and \tlabel{Goal Concurrency}
where we use Lemmas \ref{lem:mono-goals} and \ref{lem:isolationstar} to show that
concurrent derivations can always be replicated by a sequence of sequential
goal-based derivations. 
\end{proof}
 
\begin{theorem} [Sequential Reachability of Concurrent Derivations] \label{theo:corr-par-trans}
   For any initial CHR state $\sigma$, CHR state $\sigma'$ and CHR Program ${\cal P}$, 
   if $\sigma \partransstar \sigma'$ then $\sigma \goaltransstar \sigma'$.
\end{theorem}

The above follows directly from Lemma \ref{lem:corr-par-trans}
by converting each single step concurrent derivation into a sequence of sequential
derivations, and showing their composibility.


From Theorem \ref{theo:corr-goal-trans} and \ref{theo:corr-par-trans}, we have the 
following corollary, which states the correspondence between concurrent goal-based
CHR derivations and abstract CHR derivations.

\begin{corollary} [Correspondence of Concurrent Derivations] \label{corol:corr-par-trans}
   For any reachable CHR state $\chrstate{G}{Sn}$, CHR state $\chrstate{G'}{Sn'}$ and 
   CHR program ${\cal P}$, 
   \bda{ll} 
    \mbox{if }   & \chrstate{G}{Sn} \partransstar \chrstate{G'}{Sn'} \\
    \mbox{then } & (NoIds(G) \uplus DropIds(Sn)) = (NoIds(G') \uplus DropIds(Sn')) \sgap \vee \\
                 & (NoIds(G) \uplus DropIds(Sn)) \abstransstar (NoIds(G') \uplus DropIds(Sn'))
   \eda
   where $NoIds = \{c \mid c \in G, c \mbox{ is a CHR constraint}\} 
          \uplus \{e \mid e \in G, e \mbox{ is an equation}\}$
\end{corollary}

\subsubsection{Correspondence of Termination}

We show that all derivations from an initial state to final states in the concurrent 
goal-based semantics correspond to some derivation from a store to a final store
in the abstract semantics. We first define rule head instances:

\begin{definition} [Rule head instances] \label{def:rule-inst}
   For any CHR state $\sigma = \langle G,Sn \rangle$ and CHR program $\cal P$, 
   any $(H_P \stcup H_S) \subseteq Sn$ is known as a rule head instance of $\sigma$, if and only if
   $\exists (r~\atsign~H_P' \backslash H_P' \simparrow t_g \mid B) \in {\cal P}$,$\exists \phi$ 
   $Eqs(Sn) \models \phi \wedge t_g$ and $\phi(H_P' \uplus H_S') = DropIds(H_P \stcup H_S)$.
\end{definition}

\begin{definition} [Active rule head instances] \label{def:act-rule-inst}
   For any CHR state $\sigma = \langle G,Sn \rangle$ and CHR program $\cal P$, a 
   rule head instance $H$ of $\sigma$ is said to be {\em active} if and only if 
   there exists at least one $c\#i \in G$ such that $c\#i \in H$.  
\end{definition}

Rule head instances (Definition \ref{def:rule-inst}) are basically minimal subsets of the store
which matches a rule head. Active rule head instance (Definition \ref{def:act-rule-inst}) additional
have at least one of it is numbered constraint $c\#i$ in the goals as well. Therefore, by the
definition of the goal-based semantics, active rule head instances will eventually be triggered
by either the \tlabel{Simplify} or \tlabel{Propagate} derivation steps.

\begin{lemma} [Rule instances in reachable states are always active] \label{lem:act-rule-inst}
  For any reachable CHR state $\chrstate{G}{Sn}$, any rule head instance $H \subseteq Sn$ must 
  be active. i.e. $\exists c\#i \in H$ such that $c\#i \in G$.
\end{lemma}

Lemma \ref{lem:act-rule-inst} shows that all rule head instances in reachable states are always
active. This means that by applying the semantics steps in any way, we must eventually apply
the rule head instances as long as all it is constraints remain in the store. 

Theorem \ref{theo:corr-terminate} states that termination of a concurrent goal-based derivation
corresponds to termination in the abstract semantics. This is of course, provided that the
CHR program is terminating.

\begin{theorem} [Correspondence of Termination] \label{theo:corr-terminate}
   For any initial CHR state $\langle G,\{\} \rangle$, final CHR state $\langle \{\},Sn \rangle$ and
   terminating CHR program $\cal P$,
   \bda{l}
     \mbox{if } \chrstate{G}{\{\}} \partransstar \chrstate{\{\}}{Sn} \\
     \mbox{then } G \abstransstar DropIds(Sn) \mbox{ and } Final_{\cal A}(DropIds(Sn))
   \eda
\end{theorem}

We prove this theorem by first using Theorem \ref{theo:corr-par-trans} which guarantees that
a concurrent goal-based derivation from an initial state to a final state corresponds to 
some abstract semantics derivation. We next show that final states correspond to final
stores in the abstract semantics. This is done by contradiction, showing that assuming
otherwise contradicts with Lemma \ref{lem:act-rule-inst}.

\subsection{Concurrent CHR Optimizations} \label{sec:concurrent-chr-opt}

In the sequential setting, there exist
a wealth of optimizations~\cite{greg:thesis,DBLP:conf/iclp/Schrijvers05,DBLP:conf/iclp/SneyersSD05}
to speed up the execution of CHR.
Fortunately, many of these methods are still largely applicable to our concurrent goal-based 
variant as we discuss in the following. For the remainder, we assume that each goal (thread) tries the CHR 
rules from top-to-bottom to match the rule execution order assumed in
~\cite{greg:thesis,DBLP:conf/iclp/Schrijvers05,DBLP:conf/iclp/SneyersSD05}.

{\em Basic constraint indexing} like lookups via hashtables are still applicable with minor adaptations. For 
instance, the underlying hashtable implementation must be thread safe. Consider the following example:
\bda{c}
  r0 \atsign A(x,y), B(x), C(y) \simparrow x > y \mid D(x,y)
\eda
Suppose we have the active constraint $A(1,2)\#n$. To search for a
partner constraint of the form $B(1)\#m$ and $C(2)\#p$, standard CHR compilation techniques would 
optimize with indexing (hashtables) which allows constant time lookup for these constraints. 
The use of such indexing techniques is clearly applicable in a concurrent goal execution setting 
as long as concurrent access of the indexing data structures are handled properly. For example,
we can possibly have a concurrent active constraint $A(1,3)\#q$ which will compete with $A(1,2)\#n$ 
for a matching partner $B(1)\#m$. As such, hashtable implementations that facilitate such indexing 
must be able to be accessed and modified concurrently. 

{\em Guard optimizations/simplifications} aim at simplifying guard constraints by replacing
guard conditions with equivalent but simplified forms. Since guards are purely declarative, 
they are not influenced by concurrently executing goal threads (i.e.~CHR rules). Hence, all existing 
guard optimizations carry over to the concurrent setting.

The join order of a CHR rule determines the order in which partner constraints are searched
to execute a rule. The standard CHR optimization known as {\em optimal join-ordering} and 
{\em early guard scheduling} \cite{greg:thesis} aims at executing goals with the most optimal order of 
partner constraints lookup and guard testing. By optimal, we refer to maximizing the use of 
constant time index lookup. Considering the same CHR rule ($r0$) above, given the 
active constraint $B(x)$, an optimal join-ordering is to lookup for $A(x,y)$, schedule guard 
$x > y$, then lookup for $C(y)$. Since our concurrent semantics does not restrict the order 
in which partner constraints are matched, optimal join ordering and early guard scheduling are still
applicable.
 

Another set of optimizations tries to minimize the search for partner constraints by skipping 
definitely failing searches. Consider the following example:
\bda{c}
 r1 \atsign A \simparrow ... \\
 r2 \atsign A, B \simparrow ...
\eda
If the active goal $A$ cannot fire rule \tlabel{r1} then we cannot fire rule \tlabel{r2} either.
Hence, after failing to fire rule \tlabel{r1} we can drop goal $A$. Thus, we optimize away
some definitely failing search.
This statement is immediately true in the sequential setting where no other thread affects the constraint store.
The situation is different in a concurrent setting
where some other thread may have added in between the missing constraint $A$.
Then, even after failing to fire \tlabel{r1} we could fire rule \tlabel{r2}.
However, we can argue that the optimization is still valid for this example.
We will not violate the important condition to execute CHR rules exhaustively because
the newly added constraint $A$ will eventually be executed by a goal thread
which then fires rule \tlabel{r1}. 
Hence, the only concern is here that the optimization leads to indeterminism
in the execution order of CHR rules which is anyway unavoidable in a concurrent setting.

Yet there are existing optimizations which are not applicable in the concurrent setting. For example,
{\em continuation optimizations} \cite{greg:thesis,DBLP:conf/iclp/Schrijvers05} are not entirely applicable. 
Consider the following CHR rule:
\bda{c}
  r4 \atsign A(x),A(y) \simparrow x == y \mid ...
\eda
Given an active constraint $A(1)\#n$, fail continuation optimization will infer that if we fail to fire the
rule with $A(1)\#n$ matching $A(x)$, there is no point trying to match it with $A(y)$ because it will most
certainly fail as well, assuming that the store never changes. In a concurrent goal execution setting, we
cannot assume that the store never changes (while trying to execute a CHR). For instance, after failing to 
trigger the rule by matching $A(1)\#n$ with $A(x)$, suppose that a new active goal $A(1)\#m$ is added to 
the store concurrently. Now when we match $A(1)\#n$ to $A(y)$ we can find match the partner $A(1)\#m$ with 
$A(x)$, hence breaking the assumptions of the fail continuation optimization.


Late (also known as delayed) storage optimization \cite{greg:thesis} aims at delaying the storage of a goal $g$, until the 
latest point of its execution where $g$ is possibly a partner constraint of another active constraint.
Consider the following example:

\bda{l}
r1 \atsign P_1 \proparrow Q \\
r2 \atsign P_2, T_1 \simparrow R \\
r3 \atsign P_3, R_1 \simparrow True \\
r4 \atsign P_4 \proparrow S \\
r5 \atsign P_5, S_1 \simparrow True
\eda

Note to distinguish the rule heads, we annotate each rule head with a subscript integer (eg. $P_x$).
With late storage analysis techniques described in \cite{greg:thesis}, we can delay storage of an active
constraint $P$ until just before the execution of the body of $r4$. This is because the execution of
goal $S$ (obtained from firing of $r4$) can possibly trigger $r5$. While this is safe in the sequential
goal execution scheme, it is possible that rule matches are missing in the concurrent goal execution
setting. Consider the case where we have some simultaneously active goals $P\#n$ and $T\#m$. Since $P\#n$ is 
only stored when its execution has reached $r4$, the match $r2$ can be missed entirely by both active 
parallel goals $P\#n$ and $T\#m$. Specifically, this happens if goal $T\#m$ is activated only after 
$P\#n$ has tried matching with $P_2$ (of $r2$), but completes goal execution (by trying $T_1$ of $r2$,
and failing to match) before goal $P\#n$ is stored. Hence, we conclude that we cannot safely implement
late storage in the concurrent setting.

\ignore{
Late storage optimization \cite{greg:thesis} which avoids unnecessary adding and deleting (administration) 
of CHR goals is not applicable in our concurrent goal-based semantics because rule head matches can be missed if
goals delay goal storage.

There are, however, optimizations which are only applicable in the sequential setting. For example,
late storage optimization which avoids unnecessary adding and deleting (administration) of CHR goals
is not applicable in our concurrent goal-based semantics because rule head matches can be missed if
goals delay goal storage.
\ms{todo, brief example explaining this point}

\ms{TODO: pl elaborate/rewrite, too dense hard to follow.}Pruning of passive rule head occurrence can be done
consistently but under a more conservative scheme based on weaker confluence behavior of the concurrent 
goal-based semantics (This is because goals are executed in parallel and rule head occurrences can be 
tried in any arbitral ordering).
\ms{
Further comments:
The issue here is (a) guard optimization and (b) continuation optimization,
see~\cite{DBLP:conf/iclp/SneyersSD05}.
Each goal (thread) tries the rules in top-to-bottom order (need to say this somewhere).
Hence, I'd expect that *all of* the existing guard and continuation optimizations for the refined semantics are still applicable.
What am I missing?
If this is true we'd need to formally verify this claim.
}

}

\section{Related Work} \label{sec:related}

We review prior work on execution schemes 
for CHR and production rule systems.

\subsection{CHR Execution Schemes}

There exists a wealth of prior work on the semantics of CHR. We refer to~\cite{chr-survey}
for a comprehensive summary. Our focus here is on the \emph{operational} CHR semantics
and we briefly review the most relevant works.

The theoretical (a.k.a.~high-level) operational semantics~\cite{fruehwirth98:chr:art}
is derived from the abstract semantics and inherits its high degree of indeterminism.
The theoretical semantics has been mainly used for the study of high-level properties such as
confluence~\cite{abdennadher:confluence,Abdennadher99confluenceand}.
Confluence analysis has been exploited to study the degree
of concurrency in CHR programs~\cite{union-find,DBLP:conf/iclp/SchrijversS08}.
None of these works however provide direct glues how to systematically
execute concurrent programs.

In~\cite{DuckSBH04,rp-chr,rp-chr} some systematic, highly deterministic semantics have been developed
to achieve efficient implementation schemes. However, these semantics are inherently single-threaded.
Our motivation is to obtain systematic yet concurrent semantics which led us to develop
the goal-based concurrent semantics presented in this paper.
In the special case of a single goal thread, our semantics is equivalent 
to the refined operational semantics given in~\cite{DuckSBH04,rp-chr,rp-chr}.

There are only few works which explore different semantics, other than the theoretical or abstract semantics, 
to address concurrency.  The work in~\cite{DBLP:conf/padl/Sarna-StarostaR07} adopts a set-based semantics
and supports tabled, possible concurrent, rule execution. 
This execution scheme is not applicable to
CHR programs in general which usually assume a multi-set based semantics.
The recent work in~\cite{betz_raiser_fru_persistent_constraints_wlp09}
takes a new stab at concurrency by introducing the notion of persistent
constraints. The idea is to split the store into linear (multi-set like) and persistent (set like)
constraints. We are not aware of any evidence which shows that this approach 
supports effective concurrency in practice. Our approach leads to
an efficient parallel implementation as we explain in the next section.

\subsection{From Concurrent to Parallel CHR Execution} \label{sec:from-concurrent-to-parallel}

In our earlier works~\cite{chr-stm,parallel-chr} we have developed a parallel CHR implementation
scheme based on an informally described concurrent goal-based execution scheme, 
see Section 3 in~\cite{parallel-chr}. 
The present works provides a concise formal treatment of the implemented concurrent goal-based 
execution scheme. In our implementation, multiple threads, each executing a unique CHR goal, 
are executed in parallel on multiple processor cores. Parallel goal executions are largely asynchronous, 
only implicitly synchronizing via the shared constraint store. Atomic CHR execution is guaranteed via 
advanced synchronization primitives such as Software Transactional Memory.
We refer to~\cite{parallel-chr} for a thorough description of the more subtle implementation details.
Our experimental results reported in~\cite{parallel-chr} show that we achieve good scalability
when the number of processor cores increases. The overhead of the parallel implementation 
is fairly minor compared to a single-threaded implementations thanks to the use of lock-free algorithms.
Optimization methods applicable in the concurrent/parallel setting
are discussed in the earlier Section~\ref{sec:concurrent-chr-opt}.

\subsection{Parallel Production Rule Systems}

Parallel execution models of forward chaining production rule based languages 
(e.g. OPS5 \cite{ForgyM77}) have been widely studied in the context of production 
rule systems. A production rule system is defined by a set of multi-headed 
production rules (analogous to CHR rules) and a set of assertions (analogous to 
the CHR store). Production rule systems are richer than the CHR language, consisting of
user definable execution strategies and negated rule heads. This makes parallelizing 
production rule execution extremely difficult, because rule application is not monotonic 
(rules may not be applied in a larger context). As such, many previous works in parallel 
production rule systems focuses on efficient means of maintaining correctness of parallel 
rule execution (e.g. data dependency analysis \cite{627435}, sequential to parallel program 
transformation \cite{98939}), with respect to such user specified execution strategies. 
These works can be classified under two approaches, namely synchronous and asynchronous 
parallel production systems.

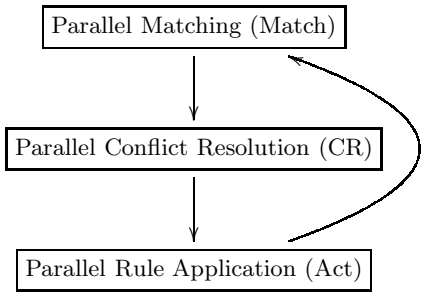
\fig{fig:production-cycle}{Parallel Production Rule Execution Cycles}{
 \xymatrix{
    \bgap \bgap \bgap \bgap \bgap 
       & \fbox{Parallel Matching (Match)} \ar[d] \\
       & \fbox{Parallel Conflict Resolution (CR)} \ar[d] \\
       & \fbox{Parallel Rule Application (Act)} \arat/_30mm/[uu] 
 }
}

For synchronous parallel production systems (e.g. UMPOPS \cite{GuptaFKNT88}), 
multiple processors/threads run in parallel. They are synchronized by execution cycles 
of the production systems. 
Figure \ref{fig:production-cycle} illustrates the 
production cycle of a typical production rule system, consisting of three execution 
phases. In the (Match) phase, all rule matches are computed. Conflict resolution (CR) 
involves filtering out matches that do not conform to the user specified rule execution
strategy, while (Act) applies the rule matches that remains (known as the eligible 
set) after the (CR) phase. By synchronizing parallel rule execution in production
cycles, a larger class of user specified execution strategies can be supported
since execution is staged.

Matching in synchronous production rule systems often use some variant of the 
RETE network \cite{Forgy82}. RETE is an incremental matching algorithm where matching 
is done eagerly (data driven) in that each newly added assertion (constraint in CHR context) 
triggers computation of all it is possible matches to rule heads. Figure \ref{fig:chr-rete} 
illustrates a RETE network (acyclic graph), described in CHR context. Root node is the 
entrance where new constraints are added. Intermediate nodes with single output edges are 
known as alpha nodes. Intermediate nodes with two output edges are beta nodes, representing 
joins between alpha nodes. Each alpha node is associated with a set of constraint matching 
its pattern, while a beta node is associated with a set of partial/complete matches. Parallel 
implementation of RETE \cite{85035} allows distinct parts of the network to be computed in parallel.
\comment{
   \xymatrix{
     & & \fbox{Entrance} \ar[dll] \ar[d] \ar[drr] & & \\
    \fbox{A(x) $M_1$} \ar[dr] & & \fbox{B(x) $M_2$} \ar[dl] & & \fbox{C(y) $M_3$} \ar[ddl] \\ 
    & \fbox{A(x),B(x) $M_4$} \ar[drr] & & & \\
    & & & \fbox{A(x),B(x),C(y) $M_5$} \ar[d] & \\
    & & & r1 & 
   }
}
\fig{fig:chr-rete}{Example of a RETE network, in CHR context}{
   \bda{c}
      r1~\atsign~A(x) \backslash B(x),C(y) \simparrow D(x,y) \bgap \{A(1),A(2),B(1),B(2),C(3)\}
   \eda
   \hspace{7mm}
   \xymatrix{
    & \fbox{Entrance} \ar[dl] \ar[d] \ar[dr] & \\
    \fbox{A(x) $M_1$} \ar[d] & \fbox{B(x) $M_2$} \ar[dl] & \fbox{C(y) $M_3$} \ar[ddl] \\ 
    \fbox{A(x),B(x) $M_4$} \ar[dr] & & \\
    & \fbox{A(x),B(x),C(y) $M_5$} \ar[d] & \\
    & r1 & 
   }
   \bda{ll}
      \bgap & M_1 = \{A(1),A(2)\} \sgap M_2 = \{B(1),B(2)\} \sgap M_3 = \{C(3)\} \\
            & M_4 = \{\{A(1),B(1)\},\{A(2),B(2)\}\} \\
            & M_5 = \{\{A(1),B(1),C(3)\},\{A(2),B(2),C(3)\}\}
   \eda
}

The most distinct characteristic of RETE is that partial matches are computed and stored.
This and the eager nature of RETE matching is suitable for production rule systems
as assertions (constraints) are propagated (not deleted) by default. Hence computing
all matches rarely results in redundancy. Traditional CHR systems do not advocate
this eager matching scheme because doing so results to many redundancies, due to
overlapping simplified matching heads. Eager matching algorithms is also proved in
\cite{leaps90} to have a larger asymptotic worst-case space complexity than lazy 
matching algorithms. 

In \cite{79070}, the matching algorithm TREAT is proposed. TREAT is similar to RETE,
except it does not store partial matches. TREAT performs better than RETE if the
overhead of maintaining and storing partial matches outweighs that of re-computing
partial matches.

Asynchronous parallel production rule systems (e.g. Swarm \cite{98939}, CREL \cite{898995}) 
introduce parallel rule execution via asynchronously running processors/threads. In such 
systems, rules can fire asynchronously (not synchronized by production cycles), hence 
enforcing execution strategies is more difficult and limited. Similar to implementations of 
goal based CHR semantics rule matching is such systems often use a variant of the 
LEAPS \cite{leaps90} lazy matching algorithm. 

\subsubsection{Observations}

\ignore{
Prior to our previous work in \cite{chr-stm,parallel-chr}, there has been no
research into the implementation of parallel execution of CHR rewritings. As
such, while works in traditional CHR systems have concluded that lazy matching 
execution model (Section \ref{ssec:goal-execution}) is most ideal for 
single-threaded execution of CHR rewrite rules, we have re-evaluated this
decision in a parallel implementation and came to the same conclusion.
}

Staging executions in synchronous parallel production rule systems allows for
flexibility in imposing execution strategies, but at a cost. In 
\cite{Neiman91controlissues}, synchronous execution of UMPOPS production rule 
system is shown to be less efficient than asynchronous execution. Hence it
is clear that synchronous systems will only be necessary if we wish to impose
some form of execution strategies on top of the abstract CHR semantics
(e.g. rule-priority, refined operational semantics). We are interested in
concurrent CHR semantics on the abstract CHR semantics.
Its non-determinism and monotonicity property provides us with the flexibility
to avoid executing threads in strict staging cycles. Thus our approach is
very similar to asynchronous parallel production rule systems. 

Lazy matching in single-threaded CHR execution is the best choice, since we
only ever have one thread of execution and wish to avoid computing overlapping 
(redundant) rule head matches. No doubt that in a parallel setting, eager matching
(like RETE, TREAT) may be more optimal if the executed CHR program consist of rules 
with more propagated heads. This is because we compute more matches in parallel with 
brute force (find all match) parallelism and we can get away with less redundancy. 
Yet to cater for the general case (more simplified heads), we again choose
lazy matching.

We therefore conclude that the goal-based execution model of CHR is still the
ideal choice for a parallel implementation of the abstract CHR semantics.
For CHR with rule priorities or 
refined CHR operational semantics,
a variant of the synchronous parallel production rule execution
is a possible choice. We leave this topic for future work.

\comment{
\begin{itemize}
  \item Production rule systems -> richer framework than CHR. Consist of
        a complex user definable execution strategies. E.g. Lex strategy
        in OPS5. 
  \item Also allows 'negated' heads. 
  \item Combine above 2 points, production rule systems do not have the
        monotonicity property, making parallel execution non-trivial. 
	\item Synchronous Parallel Execution models
	      \begin{itemize}
	        \item E.g. OPS5, Soar
	        \item Uses synchronous production cycles. i.e. Match->Select->Act.
	              Synchronous because all threads must work in the same
	              cycle.
	        \item Quote works parallelize the Match part. Uses parallel implementation
	              of RETE or TREAT. Details on RETE matching 
	        \item RETE networks: Incrementally compute and store all (partial/complete) 
	              matches, so no re-execution. Parallel thread can update different parts 
	              of the network.
	        \item RETE uses a lot of memory because it stores partial matches.
	        \item TREAT: same as RETE, but do not store partial matches. Performs better than
	              RETE in general.
	        \item Both finds all complete matches, makes sense in production rule systems:
	              matching rule heads are propagated by default.
	        \item Briefly quote works on parallelizing Select and Act parts.
        \end{itemize}
	\item Asynchronous Parallel Execution models
        \begin{itemize}
	        \item E.g. Swarm, CREL. UMPOPS allows asynchronous rule-firing but must be explicitly
	              specified by the user.  
	        \item Very similar to goal-based CHR execution models, uses some form of lazy matching 
	              like LEAPS. Hence asynchronous, as rules can fire immediately once matched (no production
	              cycles).
	        \item Swarm language built on top of a shared data space logic programming language.
	        \item Main focus and challenge of asynchronous parallel execution models in production rule systems
	              is to derive language construct support to impose execution strategies without the synchronous
	              conflict resolution in production cycles of synchronous parallel production rule systems. Hence, 
	              focuses on different concerns from our work. 
        \end{itemize}
\end{itemize}
}

\comment{
make a case that "rule parallelism" production rule style not "suitable"

\begin{itemize}
	\item Summary table of CHR goal based execution and parallel production rules
	\item We pick goal-based CHR execution model:
        \begin{itemize}
	        \item We want to exploit CHR monotonicity. Hence asynchronous model 
        \end{itemize}
\end{itemize}
}

\section{Conclusion} \label{sec:conc}

We have introduced a novel concurrent goal-based CHR semantics which is inspired
by traditional single-threaded (sequential) goal-based CHR execution models. 
Existing CHR semantics aim at introducing specific execution strategies
(e.g. ordered goal execution, rule priorities) on top of the CHR abstract 
semantics, hence adding more determinism. In contrast, the concurrent goal-based 
CHR semantics exploits the inherent non-deterministic and concurrent abstract CHR 
semantics, while introducing a systematic goal-based execution strategy.
We have shown that all concurrent derivations 
can be replicated in the sequential goal-based semantics and that there is a 
correspondence between the sequential goal-based semantics and the abstract CHR 
semantics. Thus, establishing correctness of our concurrent goal-based CHR semantics. 
Our semantics provides the basis for an efficient parallel CHR implementation.
The details are studied elsewhere~\cite{parallel-chr}.

An interesting question is how our concurrent semantics can help to
parallelize an existing single-threaded semantics such as~\cite{DuckSBH04}.
We leave the study of this issue for future work.

\section*{Acknowledgments}

We thank the reviewers for their helpful comments
on a previous version of this paper.

\section{Proofs} \label{sec:proofs}

\fig{fig:k-closures}{$k$-closure derivation steps}{
  \mbox{\bf Sequential Goal-based Semantics $k$-closure}
  \bda{c}
    \tlabel{k-Step} \bgap \bgap
    \sigma \goaltrans^0 \sigma \bgap \bgap
    \myirule{\sigma  \goaltranssf{\delta} \sigma' \sgap
             \sigma' \goaltrans^k \sigma''}{\sigma \goaltrans^{k+1} \sigma''}
  \eda
  \mbox{\bf Concurrent Goal-based Semantics $k$-closure}
  \bda{c}
    \tlabel{k-Step} \bgap \bgap
    \sigma \partrans^0 \sigma \bgap \bgap
    \myirule{\sigma  \partranssf{\delta} \sigma' \sgap
             \sigma' \partrans^k \sigma''}{\sigma \partrans^{k+1} \sigma''}
  \eda
}

In this section, we provide the proofs of the Lemmas and Theorems discussed in this 
paper. Because many of our proofs rely on inductive steps on the derivations, 
we define $k$-step derivations to facilitate the proof mechanisms. Figure 
\ref{fig:k-closures} shows $k$-step derivations of the sequential goal-based derivations 
$\goaltranssf{\delta}$ and the concurrent goal-based derivations $\partranssf{\delta}$. 

\subsubsection{Proof of Correspondence of Derivations}

\paragraph{\bf Theorem \ref{theo:corr-goal-trans} (Correspondence of Sequential Derivations)}

For any reachable CHR state $\chrstate{G}{Sn}$, CHR state $\chrstate{G'}{Sn'}$ and 
CHR Program ${\cal P}$, 
\bda{ll} 
 \mbox{if }   & \chrstate{G}{Sn} \goaltransstar \chrstate{G'}{Sn'} \\
 \mbox{then } & (NoIds(G) \uplus DropIds(Sn)) = (NoIds(G') \uplus DropIds(Sn')) \sgap \vee \\
              & (NoIds(G) \uplus DropIds(Sn)) \abstransstar (NoIds(G') \uplus DropIds(Sn'))
\eda
where $NoIds = \{c \mid c \in G, c \mbox{ is a CHR constraint}\} 
        \uplus \{e \mid e \in G, e \mbox{ is an equation}\}$

\begin{proof}
  We prove that for all finite $n$ and reachable states $\chrstate{G}{Sn},\chrstate{G'}{Sn'}$, 
  $\chrstate{G}{Sn} \goaltrans^n \chrstate{G'}{Sn'}$ either yields equivalent abstract stores or 
  corresponds to some abstract semantics derivation. We prove by induction on the derivation steps $n$. 
  Showing that goal-based derivation of any finite $n$ steps satisfying one of the following conditions: 
  
  \begin{itemize}
	  \item {\bf (C1)} $(NoIds(G) \uplus DropIds(Sn)) = (NoIds(G') \uplus DropIds(Sn'))$
	  \item {\bf (C2)} $(NoIds(G) \uplus DropIds(Sn)) \abstransstar (NoIds(G') \uplus DropIds(Sn'))$
  \end{itemize}
  
  We have the following axioms, by definition of the functions $NoIds$ and $DropIds$, for any goals $G$
  or store $Sn$:
  
  \begin{itemize}
	  \item {\bf (a1)} For any equation $e$, $NoIds(\{e\} \uplus G) = \{e\} \uplus NoIds(G)$
	  \item {\bf (a2)} For any equation $e$, $DropIds(\{e\} \stcup Sn) = \{e\} \uplus DropIds(Sn)$
	  \item {\bf (a3)} For any numbered constraint $c\#i$, $NoIds(\{c\#i\} \uplus G) = NoIds(G)$
	  \item {\bf (a4)} For any numbered constraint $c\#i$, $DropIds(\{c\#i\} \stcup Sn) = \{c\} \uplus DropIds(Sn)$
	  \item {\bf (a5)} For any CHR constraint $c$, $NoIds(\{c\} \uplus G) = \{c\} \uplus NoIds(G)$ 
	  \item {\bf (a6)} For any store $Sn'$, $DropIds(Sn \stcup Sn') = DropIds(Sn) \uplus DropIds(Sn')$
  \end{itemize}
  
  $(a1)$ and $(a2)$ are so because $NoIds$ and $DropIds$ have no effect on equations. $(a3)$ is true
  because $NoIds$ is defined to drop numbered constraints. $(a4)$ is true because $DropIds$ is
  defined to remove identifier components of numbered constraints. We have $(a5)$ because $NoIds$ has
  no effect on CHR constraints. By definition of $DropIds$, $(a6)$ is true.
  
  {\bf Base case:} We consider $\chrstate{G}{Sn} \goaltrans^0 \chrstate{G'}{Sn'}$. By
  definition of $\goaltrans^0$, we have $G = G'$ and $Sn = Sn'$. Hence 
  $(NoIds(G) \uplus DropIds(Sn)) = (NoIds(G') \uplus DropIds(Sn'))$ and we are done.
  
  {\bf Inductive case:} We assume that the theorem is true for some finite $k > 0$,
  hence $\chrstate{G}{Sn} \goaltrans^k \chrstate{G'}{Sn'}$ have some correspondence
  with the abstract semantics. 
  
  We now prove that by extending these $k$ derivations with another step, we
  preserve correspondence, namely
  $\chrstate{G}{Sn} \goaltrans^k \chrstate{G'}{Sn'} \goaltranssf{\delta} \chrstate{G''}{Sn''}$
  has a correspondence with the abstract semantics. We prove this by considering all
  possible form of derivation step, step $k+1$ can take:
  
  \begin{itemize}
	  \item \tlabel{Solve} $k+1$ step is of the form 
	       $\chrstate{\{e\} \uplus G'''}{Sn'} \goaltranssf{\delta} \chrstate{W \uplus G'''}{\{e\} \stcup Sn'}$
	       such that for some $G'''$ and $W$
	       \bda{c}
	         G' = \{e\} \uplus G''', G'' = W \uplus G''' \mbox{ and } Sn'' = \{e\} \stcup Sn' \sgap {\bf (a_{solve})}
	       \eda
	       where $e$ is an equation, $W = WakeUp(e,Sn)$ contains only goals of the form $c\#i$. This is because \tlabel{Solve} only
	       wakes up stored numbered constraints. Hence,
	       \bda{cc} 
            \hspace{-1.5cm} & 
	       \ba{lll}
	         NoIds(G'') \uplus DropIds(Sn'') & = & NoIds(W \uplus G''') \uplus DropIds(\{e\} \stcup Sn') ~~~~ {\bf (a_{solve})} \\
	                                           & = & NoIds(G''') \uplus DropIds(\{e\} \stcup Sn') ~~~~ {\bf (a3)} \\
	                                           & = & NoIds(G''') \uplus \{e\} \uplus DropIds(Sn') ~~~~ {\bf (a2)} \\
	                                           & = & NoIds(\{e\} \uplus G''') \uplus DropIds(Sn') ~~~~ {\bf (a1)} \\
	                                           & = & NoIds(G') \uplus DropIds(Sn')                ~~~~ {\bf (a_{solve})}
	       \ea
               \eda
	       Hence we can conclude that the evaluated store of derivation step $k+1$ is equivalent to abstract store of 
	       evaluated store of step $k$, therefore satisfying condition ${\bf (C1)}$.
	  \item \tlabel{Activate} $k+1$ step is of the form
	       $\chrstate{\{c\} \uplus G'''}{Sn'} \goaltranssf{\delta} \chrstate{\{c\#i\} \uplus G'''}{\{c\#i\} \stcup Sn'}$
	       such that for some $G'''$
	       \bda{c}
	         G' = \{c\} \uplus G''', G'' = \{c\#i\} \uplus G''' \mbox{ and } Sn'' = \{c\#i\} \stcup Sn' \sgap {\bf (a_{act})}
	       \eda
	       Hence,
	       \bda{cc}
                \hspace{-1.5cm} & 
	       \ba{l}
	         NoIds(G'') \uplus DropIds(Sn'') \\
                 \ba{ccll} 
                  \bgap & = & NoIds(\{c\#i\} \uplus G''') \uplus DropIds(\{c\#i\} \stcup Sn') & {\bf (a_{act})} \\
	                & = & NoIds(G''') \uplus DropIds(\{c\#i\} \stcup Sn') & {\bf (a3)} \\
	                & = & NoIds(G''') \uplus \{c\} \uplus DropIds(Sn') & {\bf (a4)} \\
	                & = & NoIds(\{c\} \uplus G''') \uplus DropIds(Sn') & {\bf (a5)} \\
	                & = & NoIds(G') \uplus DropIds(Sn')                & {\bf (a_{act})}
                 \ea
	       \ea
               \eda       
	       Hence we can conclude that evaluated store of derivation step $k+1$ is equivalent to abstract store of 
	       evaluated store of step $k$, therefore satisfying condition ${\bf (C1)}$.
	  \item \tlabel{Simplify} $k+1$ step is of the form
	        $\chrstate{\{c\#i\} \uplus G'''}{H_P \stcup \{c\#i\} \stcup H_S \stcup Sn'''} 
	         \goaltranssf{\delta} \chrstate{B \uplus G'''}{H_P \stcup Sn'''}$
	        for some $H_P$,$H_S$ and $B$ such that for some $G'''$ and $Sn'''$
	        \bda{lr}
	            Sn' = H_P \stcup \{c\#i\} \stcup H_S \stcup Sn''', Sn'' = H_P \stcup Sn''', & \\
	            G' = \{c\#i\} \uplus G''' \mbox{ and } G'' = B \uplus G''' & \sgap {\bf (a1_{simp})}
	        \eda
	        and there exists a CHR rule $r ~\atsign~ H_P' \backslash H_S' \simparrow t_g \mid B'$ 
	        such that exists $\phi$ where
	        \bda{lr}
	           DropIds(\{c\#i\} \stcup H_S) = \phi(H_S') \sgap DropIds(H_P) = \phi(H_P') & \\
	           Eq(Sn''') \models \phi \wedge t_g \sgap B = \phi(B') & {\bf (a2_{simp})}
	        \eda
	        Hence,
	        {\small
                \bda{ll}
                \hspace{-15mm} &
	        \ba{ll}
	          NoId(G') \uplus DropIds(Sn') \\
	          \bgap = NoIds(\{c\#i\} \uplus G''') \uplus DropIds(H_P \stcup \{c\#i\} \stcup H_S \stcup Sn''') & {\bf (a1_{simp})} \\
	          \bgap = NoIds(G''') \uplus DropIds(H_P \stcup \{c\#i\} \stcup H_S \stcup Sn''') & {\bf (a_3)} \\
	          \bgap = NoIds(G''') \uplus DropIds(H_P) \uplus DropIds(\{c\#i\} \stcup H_S) \uplus DropIds(Sn''') & {\bf (a_6)} \\
	          \bgap = NoIds(G''') \uplus \phi(H_P') \uplus \phi(H_S') \uplus DropIds(Sn''') & {\bf (a2_{simp})}
	        \ea
                \eda
                }
	        By definition of the abstract semantics and $a2_{simp}$, we know that we have the rule application
	        $\phi(H_P') \stcup \phi(H_S') \abstrans \phi(B')$ Therefore, by monotonicity of CHR rewriting 
	        (Theorem \ref{theo:monotonicity})
	        {\small
	        \bda{l}
	           NoId(G') \uplus DropIds(Sn') \\ 
                   \ba{lcll}
                     \bgap & = & NoIds(G''') \uplus \phi(H_P') \uplus \phi(H_S') \uplus DropIds(Sn''') \\
	                   & \abstrans & NoIds(G''') \uplus \phi(B') \uplus DropIds(Sn''') & {\bf (Theorem~\ref{theo:monotonicity})} \\
	                   & = & NoIds(\phi(B') \uplus G''') \uplus DropIds(Sn''') & {\bf (a1),(a3)} \\
	                   & = & NoIds(G'') \uplus DropIds(Sn'') & {\bf (a1_{simp})}
                   \ea
	        \eda}
	       Therefore, we have $NoId(G) \uplus DropIds(Sn)$ $\abstransstar$ $NoId(G') \uplus DropIds(Sn')$ 
               $\abstrans$ $NoIds(G'') \uplus DropIds(Sn'')$,
	       such that the $k+1$ goal-based derivation step satisfy condition ${\bf (C2)}$.  
	  \item \tlabel{Propagate} $k+1$ step is of the form
	        $\chrstate{\{c\#i\} \uplus G'''}{H_P \stcup \{c\#i\} \stcup H_S \stcup Sn'''} 
	         \goaltranssf{\delta} \chrstate{B \uplus \{c\#i\} \uplus G'''}{H_P \stcup \{c\#i\} \stcup Sn'''}$
	        for some $H_P$,$H_S$ and $B$ such that for some $G'''$ and $Sn'''$
	        \bda{lr}
	            Sn' = H_P \stcup \{c\#i\} \stcup H_S \stcup Sn''', Sn'' = H_P \stcup \{c\#i\} \stcup Sn''', & \\
	            G' = \{c\#i\} \uplus G''' \mbox{ and } G'' = B \uplus \{c\#i\} \uplus G''' & \sgap {\bf (a1_{prop})}
	        \eda
	        and there exists a CHR rule $r ~\atsign~ H_P' \backslash H_S' \simparrow t_g \mid B'$ 
	        such that exists $\phi$ where
	        \bda{lr}
	           DropIds(H_S) = \phi(H_S') \sgap DropIds(\{c\#i\} \stcup H_P) = \phi(H_P') & \\
	           Eq(Sn''') \models \phi \wedge t_g \sgap B = \phi(B') & {\bf (a2_{prop})}
	        \eda
	        Hence,
	        {\small
	        \bda{ll}
                  \hspace{-15mm} &
                  \ba{ll}
	          NoId(G') \uplus DropIds(Sn') \\
	          \bgap = NoIds(\{c\#i\} \uplus G''') \uplus DropIds(H_P \stcup \{c\#i\} \stcup H_S \stcup Sn''') & {\bf (a1_{prop})} \\
	          \bgap = NoIds(G''') \uplus DropIds(H_P \stcup \{c\#i\} \stcup H_S \stcup Sn''') & {\bf (a_3)} \\
	          \bgap = NoIds(G''') \uplus DropIds(\{c\#i\} \stcup H_P) \uplus DropIds(H_S) \uplus DropIds(Sn''') & {\bf (a_6)} \\
	          \bgap = NoIds(G''') \uplus \phi(H_P') \uplus \phi(H_S') \uplus DropIds(Sn''') & {\bf (a2_{prop})}
                  \ea
	        \eda}
	        By definition of the abstract semantics and $a2_{simp}$, we know that we have the rule application
	        $\phi(H_P') \stcup \phi(H_S') \abstrans \phi(B')$ Therefore, by monotonicity of CHR rewriting 
	        (Theorem \ref{theo:monotonicity})
	        {\small
	        \bda{l}
	           NoId(G') \uplus DropIds(Sn') \\
                   \ba{lcll} 
                    \bgap & = & NoIds(G''') \uplus \phi(H_P') \uplus \phi(H_S') \uplus DropIds(Sn''') \\
	                  & \abstrans & NoIds(G''') \uplus \phi(B') \uplus DropIds(Sn''') & {\bf (Theorem~\ref{theo:monotonicity})} \\
	                  & = & NoIds(\phi(B') \uplus G''') \uplus DropIds(Sn''') & {\bf (a1),(a5)} \\
                          & = & NoIds(\phi(B') \uplus \{c\#i\} \uplus G''') \uplus DropIds(Sn''') & {\bf (a3)} \\
	                  & = & NoIds(G'') \uplus DropIds(Sn'') & {\bf (a1_{prop})}
                   \ea
	        \eda}
	       Therefore, we have $NoId(G) \uplus DropIds(Sn)$ $\abstransstar$ $NoId(G') \uplus DropIds(Sn')$ $\abstrans$ 
               $NoIds(G'') \uplus DropIds(Sn'')$, such that the $k+1$ goal-based derivation step satisfy condition ${\bf (C2)}$.  
	  \item \tlabel{Drop} $k+1$ step is of the form
	       $\chrstate{\{c\#i\} \uplus G''}{Sn'} \goaltranssf{\delta} \chrstate{\{G''}{Sn'}$
	       such that for some $G'''$
	       \bda{c}
	         G'' = \{c\#i\} \uplus G' \mbox{ and } Sn' = Sn'' \sgap {\bf (a_{drop})}
	       \eda
	       Hence,
               \bda{ll}
               \hspace{-10mm} &
	       \ba{ccll}
	         NoIds(G'') \uplus DropIds(Sn'') & = & NoIds(\{c\#i\} \uplus G') \uplus DropIds(Sn') & {\bf (a_{drop})} \\
	                                         & = & NoIds(G') \uplus DropIds(Sn') & {\bf (a3)} 
	       \ea
               \eda	       
	       Hence we can conclude that evaluated store of derivation step $k+1$ is equivalent to abstract store of 
	       evaluated store of step $k$, therefore satisfying condition ${\bf (C1)}$.
  \end{itemize}
 
Considering all forms of $k+1$ derivation steps, \tlabel{Solve}, \tlabel{Activate} and \tlabel{Drop} satisfies
condition ${bf (C1)}$, while \tlabel{Simplify} and \tlabel{Propagate} satisfy condition ${\bf (C2)}$. Hence we 
can conclude that Theorem \ref{theo:corr-goal-trans} holds. 
\end{proof}

\paragraph{\bf Lemma \ref{lem:k-concurrency} ($k$-Concurrency)}

For any finite $k$ of mutually non-overlapping concurrent derivations,
{\small
\bda{ll}
\hspace{-10mm} &
\ba{c}
  \myirule
  {\ba{c}
   \chrstate{G_1}{H_{S1} \stcup .. \stcup H_{Si} \stcup .. \stcup H_{Sk} \stcup S}
   \partranssf{H_{P1} \backslash H_{S1}}
   \chrstate{G_1'}{\{\} \stcup .. \stcup H_{Si} \stcup .. \stcup H_{Sk} \stcup S} \\
   .. \\
   \chrstate{G_i}{H_{S1} \stcup .. \stcup H_{Si} \stcup .. \stcup H_{Sk} \stcup S}
   \partranssf{H_{Pi} \backslash H_{Si}}
   \chrstate{G_i'}{H_{S1} \stcup .. \stcup \{\} \stcup .. \stcup H_{Sk} \stcup S} \\
   .. \\
   \chrstate{G_k}{H_{S1} \stcup .. \stcup H_{Si} \stcup .. \stcup H_{Sk} \stcup S}   
   \partranssf{H_{Pk} \backslash H_{Sk}}
   \chrstate{G_k'}{H_{S1} \stcup .. \stcup H_{Si} \stcup .. \stcup \{\} \stcup S} \\
   H_{P1} \subseteq S .. H_{Pi} \subseteq S .. H_{Pk} \subseteq S \\
   \delta = H_{P1} \cup .. \cup H_{Pi} \cup .. \cup H_{Pk} \backslash
            H_{S1} \cup .. \cup H_{Si} \cup .. \cup H_{Sk}
   \ea}
  {\ba{ll}
      & \chrstate{G_1 \uplus .. \uplus G_i \uplus .. \uplus G_k \uplus G}
                 {H_{S1} \stcup .. \stcup H_{Si} \stcup .. \stcup H_{Sk} \stcup S} \\
    \partranssf{\delta}
      & \chrstate{G_1' \uplus .. \uplus G_i' \uplus .. \uplus G_k' \uplus G}
                 {S}
   \ea}
\ea
\eda
}
we can decompose this into $k-1$ applications of the (pair-wise) \tlabel{Goal Concurrency} 
derivation step.

\begin{proof}
  We prove the soundness of $k$-concurrency by showing that $k$ mutually non-overlapping 
  concurrent derivation can be decomposed into $k-1$ applications of \tlabel{Goal Concurrency} 
  step. We prove by induction on the number of concurrent derivations $k$.
  
  {\bf Base case:} $k=2$. $2$-concurrency immediately corresponds to \tlabel{Goal Concurrency}
  rule, hence it is true by definition.
  
  {\bf Inductive case:} We assume that for $j>2$ and $j<k$, we can decompose $j$ mutually 
  non-overlapping concurrent derivations. into $j-1$ applications of the \tlabel{Goal Concurrency} step. 
  We now consider $j+1$ mutually non-overlapping concurrent derivations. Because all derivations are
  non-overlapping, we can compose any two derivations amongst these $j+1$ into a single concurrent 
  step via the \tlabel{Goal Concurrency} rule. We pick any two concurrent derivations, say the $j^{th}$ 
  and $(j+1)^{th}$ (Note that by symmetry, this choice is arbitrary):
  {\small
  \bda{c}
   \chrstate{G_j}{H_{S1} \stcup .. \stcup H_{Sj} \stcup H_{Sj+1} \stcup S}
   \partranssf{H_{Pj} \backslash H_{Sj}}
   \chrstate{G_j'}{H_{S1} \stcup .. \stcup \{\} \stcup H_{Sj+1} \stcup S} \\ \\
   
   \chrstate{G_{j+1}}{H_{S1} \stcup .. \stcup H_{Sj} \stcup H_{Sj+1} \stcup S}
   \partranssf{H_{Pj+1} \backslash H_{Sj+1}}
   \chrstate{G_{j+1}'}{H_{S1} \stcup .. \stcup H_{Sj} \stcup \{\} \stcup S} \\ \\
   
   H_{Pj} \subseteq S \sgap H_{Pj+1} \subseteq S
  \eda}
  By applying the above two non-overlapping derivations with an instance of the 
  \tlabel{Goal Concurrency} rule, we have:
  {\small
   \bda{c}
    \chrstate{G_{j'}}{H_{S1} \stcup .. \stcup H_{Sj'} \stcup S}
    \partranssf{H_{Pj'} \backslash H_{Sj'}}
    \chrstate{G_{j'}'}{H_{S1} \stcup .. \stcup \{\} \stcup S} \\
    \ba{cl}
     \mbox{where} & G_{j'} = G_{j} \uplus G_{j+1} \sgap G_{j'}' = G_{j}' \uplus G_{j+1}' \\
                  & H_{Sj'} = H_{Sj} \stcup H_{Sj+1} \sgap H_{Pj'} = H_{Pj} \cup H_{Pj+1}
    \ea
   \eda 
  }
  Hence we have reduced $j+1$ non-overlapping concurrent derivations into $j$ 
  non-overlapping concurrent derivations by combining via the \tlabel{Goal Concurrency} 
  derivation step. 
  {\small
  \bda{c}
  \myirule
  {\ba{c}
   \chrstate{G_1}{H_{S1} \stcup .. \stcup H_{Sj'} \stcup S}
   \partranssf{H_{P1} \backslash H_{S1}}
   \chrstate{G_1'}{\{\} \stcup .. \stcup H_{Sj'} \stcup S} \\
   .. \\
   \chrstate{G_{j'}}{H_{S1} \stcup .. \stcup H_{Sj'} \stcup S}   
   \partranssf{H_{Pj'} \backslash H_{Sj'}}
   \chrstate{G_{j'}'}{H_{S1} \stcup .. \stcup \{\} \stcup S} \\
   H_{P1} \subseteq S .. H_{Pj'} \subseteq S \\
   \delta = H_{P1} \cup .. \cup H_{Pj'} \backslash
            H_{S1} \cup .. \cup H_{Sj'}
   \ea}
  {\ba{ll}
      & \chrstate{G_1 \uplus .. \uplus G_{j'} \uplus G}
                 {H_{S1} \stcup .. \stcup H_{Sj'} \stcup S} \\
    \partranssf{\delta}
      & \chrstate{G_1' \uplus .. \uplus G_{j'}' \uplus G}
                 {S}
   \ea}
  \eda
  }
  Hence, by our original assumption, the above is decomposable into 
  $j-1$ applications of the \tlabel{Goal Concurrency} step. This implies that
  $j+1$ concurrent derivations are decomposable into $j$ \tlabel{Goal Concurrency}
  step.
\end{proof}

\paragraph{\bf Lemma \ref{lem:mono-goals} (Monotonicity of Goals in Goal-based Semantics)}
For any goals $G$,$G'$ and $G''$ and CHR store $Sn$ and $Sn'$,
if $\chrstate{G}{Sn} \goaltransstar \chrstate{G'}{Sn'}$ 
then $\chrstate{G \uplus G''}{Sn} \goaltransstar \chrstate{G' \uplus G''}{Sn'}$

\begin{proof}
  We need to prove that for any finite $k$, if $\chrstate{G}{Sn} \goaltrans^k \chrstate{G'}{Sn'}$
  we can always extend the goals with any $G''$ such that 
  $\chrstate{G \uplus G''}{Sn} \goaltrans^k \chrstate{G' \uplus G''}{Sn'}$.

  We prove this by induction on the number of derivation steps $k$, showing that for any finite
  $i \leq k$, goals are monotonic.
  
  {\bf Base case:} We consider $\chrstate{G}{Sn} \goaltrans^0 \chrstate{G'}{Sn'}$. By
  definition of $\goaltrans^0$, we have $G = G'$ and $Sn = Sn'$. Hence we immediately
  have $\chrstate{G \uplus G''}{Sn} \goaltrans^0 \chrstate{G' \uplus G''}{Sn'}$
  
  {\bf Inductive case:} We assume that the lemma is true for some finite $i > 0$,
  hence $\chrstate{G}{Sn} \goaltrans^i \chrstate{G'}{Sn'}$ is monotonic with respect
  to the goals.
  
  We now prove that by extending these $i$ derivations with another step, 
  we still preserve monotonicity of the goals. Namely, if
  $\chrstate{G}{Sn} \goaltrans^i \chrstate{\{g\} \uplus G_i}{Sn_i} \goaltranssf{\delta} \chrstate{G_{i+1}}{Sn_{i+1}}$
  then $\chrstate{G \uplus G''}{Sn} \goaltrans^i \chrstate{G_i \uplus G''}{Sn_i} 
  \goaltranssf{\delta} \chrstate{G_{i+1} \uplus G''}{Sn_{i+1}}$
  We prove this by considering all possible form of derivation step, step $i+1^{th}$ can take:
  \begin{itemize}
	  \item \tlabel{Solve} Consider $i+1^{th}$ derivation step of the form 
	        $\chrstate{\{e\} \uplus G_i}{Sn_i} \goaltrans \chrstate{W \uplus G}{\{e\} \stcup Sn_i}$
	        for some equation $e$ and $W = WakeUp(e,Sn_i)$.
	        
	        By definition, the \tlabel{Solve} step only make reference to $e$ and $Sn_i$, hence we can
	        extend $G_i$ with any $G''$ without affecting the derivation step, i.e.
	        \bda{c}
	          \chrstate{\{e\} \uplus G_i \uplus G''}{Sn_i} \goaltrans 
	          \chrstate{W \uplus G_i \uplus G''}{\{e\} \stcup Sn_i}
	        \eda
	        Hence, given our assumption that the first $i$ derivations are monotonic with respect
          to the goals,	extending with a $i+1^{th}$ \tlabel{Solve} step preserves monotonicity
          of the goals. 
	  \item \tlabel{Activate} Consider $i+1^{th}$ derivation step of the form 
	        $\chrstate{\{c\} \uplus G_i}{Sn_i} \goaltrans \chrstate{\{c\#j\} \uplus G_i}{\{c\#j\} \stcup Sn_i}$
	        for some CHR constraint $c$, goals $G_i$ and store $Sn_i$.
	        
	        By definition, the \tlabel{Activate} step only make reference to goal $c$, hence we can
	        extend $G_i$ with any $G''$ without affecting the derivation step, i.e.
	        \bda{c}
	          \chrstate{\{c\} \uplus G_i \uplus G''}{Sn_i} \goaltrans 
	          \chrstate{\{c\#j\} \uplus G_i \uplus G''}{\{c\#j\} \stcup Sn_i}
	        \eda
	        Hence, given our assumption that the first $i$ derivations are monotonic with respect
          to the goals,	extending with a $i+1^{th}$ \tlabel{Activate} step preserves monotonicity
          of the goals. 
	  \item \tlabel{Simplify} Consider $i+1^{th}$ derivation step of the form 
	        $\chrstate{\{c\#j\} \uplus G_i}{\{c\#j\} \uplus H_S \stcup Sn_i} 
	         \goaltrans 
	         \chrstate{B \uplus G_i}{Sn_i}$
	        for some CHR constraints $H_S$ and body constraints $B$.
	        
	        By definition, the \tlabel{Simplify} step only make reference to goal $c\#j$, and $H_S$ of the
	        store, hence we can extend $G_i$ with any $G''$ without affecting the derivation step, i.e.
	        \bda{c}
	          \chrstate{\{c\#j\} \uplus G_i \uplus G''}{\{c\#j\} \stcup H_S \stcup Sn_i} \goaltrans 
	          \chrstate{B \uplus G_i \uplus G''}{Sn_i}
	        \eda
	        Hence, given our assumption that the first $i$ derivations are monotonic with respect
          to the goals,	extending with a $i+1^{th}$ \tlabel{Simplify} step preserves monotonicity
          of the goals. 
	  \item \tlabel{Propagate} Consider $i+1^{th}$ derivation step of the form 
	        $\chrstate{\{c\#j\} \uplus G_i}{H_S \stcup Sn_i} 
	         \goaltrans 
	         \chrstate{B \uplus \{c\#j\} \uplus G_i}{Sn_i}$
	        for some CHR constraints $H_S$ and body constraints $B$.
	        
	        By definition, the \tlabel{Propagate} step only make reference to goal $c\#j$, and $H_S$ of the
	        store, hence we can extend $G_i$ with any $G''$ without affecting the derivation step, i.e.
	        \bda{c}
	          \chrstate{\{c\#j\} \uplus G_i \uplus G''}{H_S \stcup Sn_i} \goaltrans 
	          \chrstate{B \uplus \{c\#j\} \uplus G_i \uplus G''}{Sn_i}
	        \eda
	        Hence, given our assumption that the first $i$ derivations are monotonic with respect
          to the goals,	extending with a $i+1^{th}$ \tlabel{Propagate} step preserves monotonicity
          of the goals. 
	  \item \tlabel{Drop} Consider $i+1^{th}$ derivation step of the form 
	        $\chrstate{\{c\#j\} \uplus G_i}{Sn_i} \goaltrans \chrstate{G_i}{Sn_i}$
	        for some numbered constraint $c\#j$.
	        
	        By definition, the \tlabel{Drop} step only make reference to goal $c\#j$, while its 
	        premise depend on $Sn_i$, hence we can extend goals $G_i$ with any $G''$ without
	        affecting the derivation step, i.e.
	        \bda{c}
	          \chrstate{\{c\#j\} \uplus G_i \uplus G''}{Sn_i} \goaltrans 
	          \chrstate{G_i \uplus G''}{Sn_i}
	        \eda
	        Hence, given our assumption that the first $i$ derivations are monotonic with respect
          to the goals,	extending with a $i+1^{th}$ \tlabel{Drop} step preserves monotonicity
          of the goals. 
  \end{itemize}
  Hence, with our assumption of monotonicity of goals for $i$ steps, the goals are still monotonic for
	$i+1$ steps regardless of the form of the $i+1^{th}$ derivation step.
\end{proof}

\paragraph{Lemma \ref{lem:isolation} (Isolation of Goal-based Derivations)} 

If $\chrstate{G}{H_P \stcup H_S \stcup S_1 \stcup S_2} 
    \goaltranssf{H_P \backslash H_S} 
    \chrstate{G'}{H_P \stcup S_1' \stcup S_2}$ 
then $\chrstate{G}{H_P \stcup H_S \stcup S_1} 
      \goaltranssf{H_P \backslash H_S} 
      \chrstate{G'}{H_P \stcup S_1'}$

\begin{proof}
  We need to show that for any goal-based derivation, we can omit any constraint
  of the store which is not a side-effect of the derivation. To prove this, we
  consider all possible forms of goal-based derivations:
  
  \begin{itemize}
	  \item \tlabel{Solve} Consider derivation of the form
	        \bda{c}
	         \chrstate{\{e\} \uplus G}{W \stcup \{\} \stcup S_1 \stcup S_2} \goaltranssf{W \backslash \{\}} 
	         \chrstate{W \uplus G}{W \stcup \{\} \stcup \{e\} \stcup S_1 \stcup S_2}
	        \eda
	        Since wake up side-effect is captured in $W$, we can drop $S_2$ without affecting the derivation.
	        Hence we also have:
	        \bda{c}
	         \chrstate{\{e\} \uplus G}{W \stcup \{\} \stcup S_1} \goaltranssf{W \backslash \{\}} 
	         \chrstate{W \uplus G}{W \stcup \{\} \stcup \{e\} \stcup S_1}
	        \eda
	  \item \tlabel{Activate} Consider derivation of the form
	        \bda{c}
	         \chrstate{\{c\} \uplus G}{\{\} \stcup \{\} \stcup S_1 \stcup S_2} \goaltranssf{\{\} \backslash \{\}} 
	         \chrstate{\{c\#i\} \uplus G}{\{\} \stcup \{\} \stcup \{c\#i\} \stcup S_1 \stcup S_2}
	        \eda
	        Since \tlabel{Activate} simply introduces a new constraint $c\#i$ into the store, we
	        can drop $S_2$ without affecting the derivation. Hence we also have:
	        \bda{c}
	         \chrstate{\{c\} \uplus G}{\{\} \stcup \{\} \stcup S_1} \goaltranssf{\{\} \backslash \{\}} 
	         \chrstate{\{c\#i\} \uplus G}{\{\} \stcup \{\} \stcup \{c\#i\} \stcup S_1}
	        \eda
	  \item \tlabel{Simplify} Consider derivation of the form
	        \bda{c}
	          \chrstate{\{c\#i\} \uplus G}{H_P \stcup H_S \stcup S_1 \stcup S_2}
	          \goaltranssf{H_P \backslash H_S}
	          \chrstate{B \uplus G}{H_P \stcup S_1 \stcup S_2}
	        \eda
	        Since $S_2$ is not part of the side-effects of this derivation, we can drop $S_2$ 
	        without affecting the derivation. Hence we also have:
	        \bda{c}
	          \chrstate{\{c\#i\} \uplus G}{H_P \stcup H_S \stcup S_1}
	          \goaltranssf{H_P \backslash H_S}
	          \chrstate{B \uplus G}{H_P \stcup S_1}
	        \eda
	  \item \tlabel{Propagate} Consider derivation of the form
	        \bda{c}
	          \chrstate{\{c\#i\} \uplus G}{H_P \stcup H_S \stcup S_1 \stcup S_2}
	          \goaltranssf{H_P \backslash H_S}
	          \chrstate{B \uplus \{c\#i\} \uplus G}{H_P \stcup S_1 \stcup S_2}
	        \eda
	        Since $S_2$ is not part of the side-effects of this derivation, we can drop $S_2$ 
	        without affecting the derivation. Hence we also have:
	        \bda{c}
	          \chrstate{\{c\#i\} \uplus G}{H_P \stcup H_S \stcup S_1}
	          \goaltranssf{H_P \backslash H_S}
	          \chrstate{B \uplus \{c\#i\} \uplus G}{H_P \stcup S_1}
	        \eda
	  \item \tlabel{Drop} Consider derivation of the form
	        \bda{c}
	         \chrstate{\{c\#i\} \uplus G}{\{\} \stcup \{\} \stcup S_1 \stcup S_2} \goaltranssf{\{\} \backslash \{\}} 
	         \chrstate{G}{\{\} \stcup \{\} \stcup S_1 \stcup S_2}
	        \eda
	        \tlabel{Drop} simply removes the goal $c\#i$ when no instances of \tlabel{Simplify} or
	        \tlabel{Propagate} can apply on it. Note that its premise references to the entire store, 
	        so removing $S_2$ may seems unsafe. But since removing constraints from the store will not 
	        cause $c\#i$ to be applicable to any instances of \tlabel{Simplify} or \tlabel{Propagate},
	        hence we also have:
	        \bda{c}
	         \chrstate{\{c\} \uplus G}{\{\} \stcup \{\} \stcup S_1} \goaltranssf{\{\} \backslash \{\}} 
	         \chrstate{G}{\{\} \stcup \{\} \stcup S_1}
	        \eda
  \end{itemize}
\end{proof}

\paragraph{Lemma \ref{lem:isolationstar} (Isolation of Transitive Goal-based Derivations)}
If $\chrstate{G}{H_P \stcup H_S \stcup S_1 \stcup S_2} 
    \goaltransstar
    \chrstate{G'}{H_P \stcup S_1' \stcup S_2}$
with side-effects $\delta = H_P \backslash H_S$, 
then $\chrstate{G}{H_P \stcup H_S \stcup S_1} 
      \goaltransstar 
      \chrstate{G'}{H_P \stcup S_1'}$

\begin{proof}
  We need to prove that for all $k$, $\chrstate{G}{H_P \stcup H_S \stcup S_1 \stcup S_2} 
  \goaltranssf{k} \chrstate{G'}{H_P \stcup S_1' \stcup S_2}$ with side-effects 
  $\delta = H_P \backslash H_S$ we can always safely omit affected potions of 
  the store from the derivation. We prove by induction on $i \leq k$.
  
  {\bf Base case:} $i=1$. Consider, $\chrstate{G}{H_P \stcup H_S \stcup S_1 \stcup S_2} 
  \goaltrans^1 \chrstate{G'}{H_P \stcup S_1' \stcup S_2}$. This corresponds
  to the premise in Lemma \ref{lem:isolation}, hence we can safely omit $S_2$
  from the derivation.
  
  {\bf Inductive case:} $i>1$. we assume that for any 
  $\chrstate{G}{H_{Pi} \stcup H_{Si} \stcup S_{1i} \stcup S_{2i}} 
  \goaltrans^i \chrstate{G'}{H_{Pi} \stcup S_{1i}' \stcup S_{2i}}$ with 
  side-effects $\delta_i = H_{Pi} \backslash H_{Si}$, we can safely omit $S_{2i}$ from the 
  derivation. Let's consider a $j = i+1$ derivation step from here, which contains side-effects 
  $\delta_j = H_{Pj} \backslash H_{Sj}$ non-overlapping with $\delta_i$. Hence $H_{Pj}$ and
  $H_{Sj}$ must be in $S_{2i}$ (i.e. $S_{2i} = H_{Pj} \stcup H_{Sj} \stcup S_{1j} \stcup S_{2j}$).
  \bda{ll}
    & \chrstate{G}{H_{Pi} \stcup H_{Si} \stcup S_{1i} \stcup H_{Pj} \stcup H_{Sj} \stcup S_{1j} \stcup S_{2j}} \\
   \goaltrans^i
    & \chrstate{G'}{H_P \stcup S_{1i}' \stcup H_{Pj} \stcup H_{Sj} \stcup S_{1j} \stcup S_{2j}} \\
   \goaltranssf{\delta_j}
    & \chrstate{G''}{H_P \stcup S_{1i}' \stcup H_{Pj} \stcup S_{1j}' \stcup S_{2j}}
  \eda
  Hence consider the following substitutions:
  \bda{lll}
     H_P = H_{Pi} \cup H_{Pj}   & \sgap & H_S = H_{Si} \cup H_{Sj} \\
     S_1 = S_{1i} \stcup S_{1j} &       & S_1' = S_{1i}' \stcup S_{1j}' \\
     \delta = H_P \backslash H_S
  \eda
  we have $\chrstate{G}{H_P \stcup H_S \stcup S_1 \stcup S_{2j}} \goaltrans^{i+1} 
  \chrstate{G}{H_P \stcup S_1' \stcup S_{2j}}$ with side-effects $\delta$ such that
  no constraints in $S_{2j}$ is in $\delta$. Hence we can safely omit $S_{2j}$ from the
  derivation and we have isolation for $i+1$ derivations as well. 
\end{proof}

\paragraph{Lemma \ref{lem:corr-par-trans} (Sequential Reachability of Concurrent Derivation Steps)} 
   
For any sequentially reachable CHR state $\sigma$, CHR state $\sigma'$ and rewriting
side-effects $\delta$ if $\sigma \partranssf{\delta} \sigma'$ then $\sigma'$ is 
sequentially reachable, $\sigma \goaltransstar \sigma'$ with side-effects $\delta$.

\begin{proof}
From the $k$-concurrency Lemma (Lemma \ref{lem:k-concurrency}) we showed that any finite 
$k$ mutually non-overlapping concurrent goal-based derivations can be replicated by nested 
application of the \tlabel{Goal Concurrency} step. Hence, to prove sequential reachability of 
concurrent derivations, we only need to consider the derivation steps \tlabel{Lift} and 
\tlabel{Goal Concurrency} which sufficiently covers the concurrent behaviour of any $k$ 
concurrent derivations. 

We prove by structural induction of the concurrent goal-based semantics derivation steps 
\tlabel{Lift} and \tlabel{Goal Concurrency}.
  \begin{itemize}
	  \item \tlabel{Lift} is the base case. Application of \tlabel{Lift} simply lifts a
	        goal-based derivation $\sigma \goaltranssf{\delta} \sigma'$ into a concurrent
	        goal-based derivation $\sigma \partranssf{\delta} \sigma'$. Thus states 
	        $\sigma'$ derived from the \tlabel{Lift} step is immediately sequentially
	        reachable since $\sigma \goaltranssf{\delta} \sigma'$ implies
	        $\sigma \goaltransstar \sigma'$.
	  \item \tlabel{Goal Concurrency}
	        {\small 
	        \bda{c}
	         \myirule{\mbox{\bf (D1)} \sgap
	             \chrstate{G_1}{H_{S1} \stcup H_{S2} \stcup S} \partranssf{\delta_1}
               \chrstate{G_1'}{\{\} \stcup H_{S2} \stcup S} \\
               \mbox{\bf (D2)} \sgap
               \chrstate{G_2}{H_{S1} \stcup H_{S2} \stcup S} \partranssf{\delta_2} 
               \chrstate{G_2'}{H_{S1} \stcup \{\} \stcup S} \\
               \delta_1 = H_{P1} \backslash H_{S1} \sgap \delta_2 = H_{P2} \backslash H_{S2} \\
               H_{P1} \subseteq S \sgap H_{P2} \subseteq S \sgap \delta = H_{P1} \stcup H_{P2} \backslash H_{S1} \stcup H_{S2}}
              {\ba{ll}
                 & \chrstate{G_1 \uplus G_2 \uplus G}{H_{S1} \stcup H_{S2} \stcup S} \\
               \mbox{\bf (C)} \sgap \partranssf{\delta} 
                 & \chrstate{G_1' \uplus G_2' \uplus G}{S}
               \ea} 
	        \eda }
	        we assume that {\bf (D1)} and {\bf (D2)} are sequentially reachable. This means that 
	        we have the following:
	        \bda{l}
	          \chrstate{G_1}{H_{S1} \stcup H_{S2} \stcup S} \goaltransstar \chrstate{G_1'}{\{\} \stcup H_{S2} \stcup S} \\
	          \mbox{ with side-effects } \delta_1 = H_{P1} \backslash H_{S1} \mbox{ such that } H_{P1} \subseteq S  \sgap {\bf (a_{D1})} \\ \\
	          \chrstate{G_2}{H_{S1} \stcup H_{S2} \stcup S} \goaltransstar \chrstate{G_2'}{H_{S1} \stcup \{\} \stcup S} \\
	          \mbox{ with side-effects } \delta_2 = H_{P2} \backslash H_{S2} \mbox{ such that } H_{P2} \subseteq S \sgap {\bf (a_{D2})}
	        \eda
	        Since both derivations are by definition non-overlapping in side-effects, we can show that {\bf (C)} is sequentially 
	        reachable, using monotonicity of goals (Lemma \ref{lem:mono-goals}) and isolation of derivations (Lemma \ref{lem:isolation}):
	        \bda{lll}
	           & \chrstate{G_1 \uplus G_2 \uplus G}{H_{S1} \stcup H_{S2} \stcup S} \\
	         \goaltransstar
	           & \chrstate{G_1' \uplus G_2 \uplus G}{H_{S2} \stcup S} & {\bf (Lemma \ref{lem:mono-goals}, a_{D1})} \\
	         \goaltransstar 
	           & \chrstate{G_1' \uplus G_2' \uplus G}{S} & {\bf (Lemma \ref{lem:mono-goals}, Lemma \ref{lem:isolationstar}, a_{D2})}
	        \eda
	        Hence, the above sequential goal-based derivation shows that \tlabel{Goal Concurrency} derivation step 
	        is sequentially reachable with side-effect $\delta$. 
  \end{itemize}
\end{proof}
 
\paragraph{\bf Theorem \ref{theo:corr-par-trans} (Sequential Reachability of Concurrent Derivations)}

For any initial CHR state $\sigma$, CHR state $\sigma'$ and CHR Program ${\cal P}$, 
if $\sigma \partransstar \sigma'$ then $\sigma \goaltransstar \sigma'$.

\begin{proof}
   We prove that for all finite $k$ number of concurrent derivation steps 
   $\sigma \partrans^k \sigma'$, we can find a corresponding sequential 
   derivation sequence $\sigma \goaltrans^* \sigma'$.
   
   {\bf Base case:} $k=1$. We consider $\sigma \partrans^1 \sigma'$. From Lemma \ref{lem:corr-par-trans},
   we can conclude that we have $\sigma \goaltrans^* \sigma'$ as well.
   
   {\bf Inductive case:} $k>1$. We consider $\sigma \partrans^k \sigma'$, assuming that it is
   sequentially reachable, hence we also have $\sigma \goaltransstar \sigma'$. We consider
   extending this derivation with the $k+1^{th}$ step $\sigma' \partrans \sigma''$. By Lemma
   \ref{lem:corr-par-trans}, we can conclude that the $k+1^{th}$ concurrent derivation is
   sequential reachable, hence $\sigma' \goaltransstar \sigma''$. Hence we have
   $\sigma \goaltransstar \sigma' \goaltransstar \sigma''$ showing that 
   $\sigma \partrans^{k+1} \sigma''$ is sequentially reachable.
\end{proof}

\subsubsection{Correspondence of Termination}

\paragraph{\bf Lemma \ref{lem:act-rule-inst} (Rule instances in reachable states are always active)}
For any reachable CHR state $\chrstate{G}{Sn}$, any rule head instance $H \subseteq Sn$ must be active.
i.e. $\exists c\#i \in H$ such that $c\#i \in G$.

\begin{proof}
   We will prove this for the sequential goal-based semantics. Since Theorem \ref{theo:corr-par-trans}
   states all concurrent derivation is sequentially reachable, this Lemma immediately applies to the
   concurrent goal-based semantics as well.
   
   We prove that for all finite $k$ derivations from any initial CHR state $\chrstate{G}{\{\}}$,
   i.e. $\chrstate{G}{\{\}} \goaltrans^k \chrstate{G'}{Sn'}$, all rule head instances $H \subseteq Sn'$
   has at least one $c\#i \in H$ such that $c\#i \in G$. We prove by induction on $i<k$ that states
   reachable by $i$ derivations from an initial stage have the above property.
   
   {\bf Base case:} $i=0$. Hence $\chrstate{G}{\{\}} \goaltrans^0 \chrstate{G'}{Sn'}$. By definition,
   $G = G'$ and $Sn' = \{\}$. Since $Sn'$ is empty, the base case immediately satisfies the Lemma.
   
   {\bf Inductive case:} $i>0$. We assume that for any $\chrstate{G}{\{\}} \goaltrans^i \chrstate{G'}{Sn'}$,
   all rule head instances $H \subseteq Sn'$ is active, hence have at least one $c\#i \in H$ such that
   $c\#i \in G'$. We extend this derivation with an $i+1^{th}$ step, hence 
   $\chrstate{G}{\{\}} \goaltrans^i \chrstate{G'}{Sn'} \goaltranssf{\delta} \chrstate{G''}{Sn''}$.
   We now prove that all rule head instances in $Sn''$ are active. We consider all possible
   forms of this $i+1^{th}$ derivation step. We omit side-effects.
   
  \begin{itemize}
	  \item \tlabel{Solve} $i+1$ derivation step is of the form 
	        $\chrstate{\{e\} \uplus G'''}{Sn'} \goaltrans \chrstate{W \uplus G'''}{\{e\} \stcup Sn'}$
	        for some goals $G'''$ and $W = WakeUp(e,Sn')$. Our assumption provides that all rule head instances
	        in $Sn'$ are active. Introducing $e$ into the store will possibly introduce new rule head instances. 
	        This is because for some CHR rule $(r~\atsign~H_P \backslash H_S \simparrow t_g\mid B) \in {\cal P}$ 
	        since we may have a new $\phi$ such that \ $Eqs(\{e\} \stcup Sn') \models \phi \wedge t_g$ and 
	        $\phi(H_P \stcup H_S) \in Sn'$. This means that there is at least one $c\#i$ in $\phi(H_P \stcup H_S)$ 
	        which is further grounded by $e$. Thankfully, by definition of $W = WakeUp(e,Sn')$, we have $c\#i \in W$. 
	        Hence new rule head instances will become active because of introduction of $W$ to the goals.
	  \item \tlabel{Activate} $i+1$ derivation step is of the form
	        $\chrstate{\{c\} \uplus G'''}{Sn'} \goaltrans \chrstate{\{c\#i\} \uplus G'''}{\{c\#i\} \stcup Sn'}$. 
	        Our assumption provides that all rule head instances in $Sn'$ are active. By adding $c\#i$ to the
	        store, we can possibly introduce new rule head instances $\{c\#i\} \stcup H$ such that $H \in Sn'$.
	        Since $c\#i$ is also retained as a goal, such new rule head instances are active as well. 
	  \item \tlabel{Simplify} $i+1$ derivation step is of the form
	        $\chrstate{\{c\#i\} \uplus G'''}{\{c\#i\} \stcup H_S \stcup Sn'} \goaltrans \chrstate{B \uplus G'''}{Sn'}$. 
	        Our assumption provides that all rule head instances in $Sn'$ are active. $c\#i$ has applied a rule
	        instance, removing $c\#i$ and some $H_S$ from the store. Since $c\#i$ is no longer in the store, we
	        can safely remove $c\#i$ from the goals. Removing $H_S$ from the store will only (possibly) remove
	        other rule head instance from the store. Hence rule head instances in $Sn'$ still remain active. 
	  \item \tlabel{Propagate} $i+1$ derivation step is of the form
	        $\chrstate{\{c\#i\} \uplus G'''}{\{c\#i\} \stcup H_S \stcup Sn'} \goaltrans 
	        \chrstate{B \uplus \{c\#i\} \uplus G'''}{\{c\#i\} \stcup Sn'}$. 
	        Our assumption provides that all rule head instances in $Sn'$ are active. $c\#i$ has applied a rule
	        instance, removing some $H_S$ from the store. Since $c\#i$ is still in the store, we
	        cannot safely remove $c\#i$ from the goals, thus it is retained. Removing $H_S$ from the store will 
	        only (possibly) remove other rule head instance from the store. Hence rule head instances in $Sn'$,
	        including those that contains $c\#i$, still remain active.
	  \item \tlabel{Drop} $i+1$ derivation step is of the form
	        $\chrstate{\{c\#i\} \uplus G'''}{Sn'} \goaltrans \chrstate{G'''}{Sn'}$. Our assumption provides that 
	        all rule head instances in $Sn'$ are active. Premise of the \tlabel{Drop} step demands that no 
	        \tlabel{Simplify} and \tlabel{Propagate} steps apply on $c\#i$. This means that $c\#i$ is not part of
	        any rule head instances in $Sn'$. Hence we can safely remove $c\#i$ from the goals without risking 
	        to deactivate any rule instances.
  \end{itemize}
  Hence \tlabel{Solve} and \tlabel{Activate} guarantees that new rule head instances become active, \tlabel{Drop}
  safely removes a goal without deactivating any rule head instances and \tlabel{Simplify} and \tlabel{Propagate}
  only removes constraint from the store. In all cases, existing rule head instances remain active while new rule
  head instances become active, thus we have proved the lemma.
\end{proof}

\paragraph{\bf Theorem \ref{theo:corr-terminate} (Correspondence of Termination)}
For any initial CHR state $\langle G,\{\} \rangle$, final CHR state $\langle \{\},Sn \rangle$ and
terminating CHR program $\cal P$,
\bda{l}
   \mbox{if } \chrstate{G}{\{\}} \partransstar \chrstate{\{\}}{Sn} \\
   \mbox{then } G \abstransstar DropIds(Sn) \mbox{ and } Final_{\cal A}(DropIds(Sn))
\eda

\begin{proof}
  We prove that for any concurrent derivation $\chrstate{G}{\{\}} \partransstar \chrstate{\{\}}{Sn}$,
  we have a corresponding abstract derivation $G \abstransstar DropIds(Sn)$. Theorem \ref{theo:corr-par-trans}
  states that we can replicate the above concurrent derivation, with a sequential derivation. Hence we have
  $\chrstate{G}{\{\}} \goaltransstar \chrstate{\{\}}{Sn}$. By instantiating Theorem \ref{theo:corr-goal-trans},
  we immediately have $G \abstransstar DropIds(Sn)$ from this sequential goal-based derivation.
  
  Next we show that $DropIds(Sn)$ is a final store ($Final_{\cal A}(DropIds(Sn))$) with respect to some
  CHR program ${\cal P}$. We prove by contradiction: Suppose $DropIds(Sn)$ is not a final store, hence 
  $\chrstate{\{\}}{Sn}$ has at least one rule head instance $H$ of ${\cal P}$ in $Sn$ which is not active,
  since the goals are empty. However, this contradicts with Lemma \ref{lem:act-rule-inst}, which states that
  all reachable states have only active rule instances. Since $\chrstate{\{\}}{Sn}$ is sequentially
  reachable, it must be the case that $Sn$ has no rule head instances of ${\cal P}$. Therefore $DropIds(Sn)$
  must be a final store.
\end{proof}


\begin{thebibliography}{}

\bibitem[\protect\citeauthoryear{Abdennadher}{Abdennadher}{1997}]{abdennadher:%
confluence}
{\sc Abdennadher, S.} 1997.
\newblock Operational semantics and confluence of constraint propagation rules.
\newblock In {\em Proc.\ of CP'97}. LNCS. Springer-Verlag, 252--266.

\bibitem[\protect\citeauthoryear{Abdennadher, Fruhwirth, and Meuss}{Abdennadher
  et~al\mbox{.}}{1999}]{Abdennadher99confluenceand}
{\sc Abdennadher, S.}, {\sc Fruhwirth, T.}, {\sc and} {\sc Meuss, H.} 1999.
\newblock Confluence and semantics of constraint simplification rules.
\newblock {\em Constraints Journal\/}~{\em 4}.

\bibitem[\protect\citeauthoryear{Betz, Raiser, and Fr{\"u}hwirth}{Betz
  et~al\mbox{.}}{2009}]{betz_raiser_fru_persistent_constraints_wlp09}
{\sc Betz, H.}, {\sc Raiser, F.}, {\sc and} {\sc Fr{\"u}hwirth, T.} 2009.
\newblock Persistent constraints in {C}onstraint {H}andling {R}ules.
\newblock In {\em WLP '09: Proc. 23rd Workshop on (Constraint) Logic
  Programming}.
\newblock To appear.

\bibitem[\protect\citeauthoryear{{De Koninck}, Stuckey, and Duck}{{De Koninck}
  et~al\mbox{.}}{2008}]{rp-chr}
{\sc {De Koninck}, L.}, {\sc Stuckey, P.}, {\sc and} {\sc Duck, G.} 2008.
\newblock Optimizing compilation of {CHR} with rule priorities.
\newblock In {\em Proc.\ of FLOPS'08}. LNCS, vol. 4989. Springer-Verlag,
  32--47.

\bibitem[\protect\citeauthoryear{D.P.~Miranker and Gadbois}{D.P.~Miranker and
  Gadbois}{1990}]{leaps90}
{\sc D.P.~Miranker, D.~Brant, B.~L.} {\sc and} {\sc Gadbois, D.} 1990.
\newblock On the performance of lazy matching in production systems.
\newblock In {\em In proceedings of International Conference on Artificial
  Intelligence AAAI}. 685--692.

\bibitem[\protect\citeauthoryear{Duck}{Duck}{2005}]{greg:thesis}
{\sc Duck, G.~J.} 2005.
\newblock Compilation of {Constraint Handling Rules}.
\newblock Ph.D. thesis, The University of Melbourne.

\bibitem[\protect\citeauthoryear{Duck, Stuckey, de~la Banda, and Holzbaur}{Duck
  et~al\mbox{.}}{2004}]{DuckSBH04}
{\sc Duck, G.~J.}, {\sc Stuckey, P.~J.}, {\sc de~la Banda, M. J.~G.}, {\sc and}
  {\sc Holzbaur, C.} 2004.
\newblock The refined operational semantics of {Constraint Handling Rules}.
\newblock In {\em Proc of ICLP'04}. LNCS, vol. 3132. Springer-Verlag, 90--104.

\bibitem[\protect\citeauthoryear{Forgy}{Forgy}{1982}]{Forgy82}
{\sc Forgy, C.} 1982.
\newblock Rete: A fast algorithm for the many patterns/many objects match
  problem.
\newblock {\em Artif. Intell.\/}~{\em 19,\/}~1, 17--37.

\bibitem[\protect\citeauthoryear{Forgy and McDermott}{Forgy and
  McDermott}{1977}]{ForgyM77}
{\sc Forgy, C.} {\sc and} {\sc McDermott, J.~P.} 1977.
\newblock Ops, a domain-independent production system language.
\newblock In {\em IJCAI}. 933--939.

\bibitem[\protect\citeauthoryear{Fr{\"u}hwirth}{Fr{\"u}hwirth}{1998}]{fruehwir%
th98:chr:art}
{\sc Fr{\"u}hwirth, T.} 1998.
\newblock Theory and practice of constraint handling rules.
\newblock {\em Journal of Logic Programming, Special Issue on Constraint Logic
  Programming\/}~{\em 37,\/}~1-3, 95--138.

\bibitem[\protect\citeauthoryear{Fr{\"u}hwirth}{Fr{\"u}hwirth}{2005}]{union-fi%
nd}
{\sc Fr{\"u}hwirth, T.} 2005.
\newblock Parallelizing union-find in {Constraint Handling Rules} using
  confluence analysis.
\newblock In {\em Proc.\ of ICLP'05}. LNCS, vol. 3668. Springer-Verlag,
  113--127.

\bibitem[\protect\citeauthoryear{Fr{\"u}hwirth}{Fr{\"u}hwirth}{2006}]{1140337}
{\sc Fr{\"u}hwirth, T.} 2006.
\newblock Constraint handling rules: the story so far.
\newblock In {\em Proc.\ of PPDP '06}. ACM Press, 13--14.

\bibitem[\protect\citeauthoryear{Gamble}{Gamble}{1990}]{98939}
{\sc Gamble, R.~F.} 1990.
\newblock Transforming rule-based programs: from the sequential to the
  parallel.
\newblock In {\em IEA/AIE '90: Proceedings of the 3rd international conference
  on Industrial and engineering applications of artificial intelligence and
  expert systems}. ACM, New York, NY, USA, 854--863.

\bibitem[\protect\citeauthoryear{Gupta, Forgy, Kalp, Newell, and Tambe}{Gupta
  et~al\mbox{.}}{1988}]{GuptaFKNT88}
{\sc Gupta, A.}, {\sc Forgy, C.}, {\sc Kalp, D.}, {\sc Newell, A.}, {\sc and}
  {\sc Tambe, M.} 1988.
\newblock Parallel ops5 on the encore multimax.
\newblock In {\em ICPP (1)}. 71--280.

\bibitem[\protect\citeauthoryear{Ishida}{Ishida}{1991}]{627435}
{\sc Ishida, T.} 1991.
\newblock Parallel rule firing in production systems.
\newblock {\em IEEE Transactions on Knowledge and Data Engineering\/}~{\em
  3,\/}~1, 11--17.

\bibitem[\protect\citeauthoryear{Lam and Sulzmann}{Lam and
  Sulzmann}{2007}]{chr-stm}
{\sc Lam, E. S.~L.} {\sc and} {\sc Sulzmann, M.} 2007.
\newblock A concurrent {Constraint Handling Rules} implementation in {Haskell}
  with software transactional memory.
\newblock In {\em Proc.\ of ACM SIGPLAN Workshop on Declarative Aspects of
  Multicore Programming (DAMP'07)}. 19--24.

\bibitem[\protect\citeauthoryear{Mahajan and Kumar}{Mahajan and
  Kumar}{1990}]{85035}
{\sc Mahajan, M.} {\sc and} {\sc Kumar, V. K.~P.} 1990.
\newblock Efficient parallel implementation of rete pattern matching.
\newblock {\em Comput. Syst. Sci. Eng.\/}~{\em 5,\/}~3, 187--192.

\bibitem[\protect\citeauthoryear{Miranker}{Miranker}{1990}]{79070}
{\sc Miranker, D.~P.} 1990.
\newblock {\em TREAT: a new and efficient match algorithm for AI production
  systems}.
\newblock Morgan Kaufmann Publishers Inc., San Francisco, CA, USA.

\bibitem[\protect\citeauthoryear{Miranker, Kuo, and Browne}{Miranker
  et~al\mbox{.}}{1989}]{898995}
{\sc Miranker, D.~P.}, {\sc Kuo, C.}, {\sc and} {\sc Browne, J.~C.} 1989.
\newblock Parallelizing transformations for a concurrent rule execution
  language.
\newblock Tech. rep., Austin, TX, USA.

\bibitem[\protect\citeauthoryear{Neiman}{Neiman}{1991}]{Neiman91controlissues}
{\sc Neiman, D.~E.} 1991.
\newblock Control issues in parallel rule-firing production systems.
\newblock In {\em in Proceedings of National Conference on Artificial
  Intelligence}. 310--316.

\bibitem[\protect\citeauthoryear{Sarna-Starosta and
  Ramakrishnan}{Sarna-Starosta and
  Ramakrishnan}{2007}]{DBLP:conf/padl/Sarna-StarostaR07}
{\sc Sarna-Starosta, B.} {\sc and} {\sc Ramakrishnan, C.~R.} 2007.
\newblock Compiling constraint handling rules for efficient tabled evaluation.
\newblock In {\em Proc.\ of PADL'07}. LNCS, vol. 4354. Springer, 170--184.

\bibitem[\protect\citeauthoryear{Schrijvers}{Schrijvers}{2005}]{DBLP:conf/iclp%
/Schrijvers05}
{\sc Schrijvers, T.} 2005.
\newblock Analyses, optimizations and extensions of {Constraint Handling Rules:
  Ph.D.} summary.
\newblock In {\em Proc.\ of ICLP'05}. LNCS, vol. 3668. Springer-Verlag,
  435--436.

\bibitem[\protect\citeauthoryear{Schrijvers and Sulzmann}{Schrijvers and
  Sulzmann}{2008}]{DBLP:conf/iclp/SchrijversS08}
{\sc Schrijvers, T.} {\sc and} {\sc Sulzmann, M.} 2008.
\newblock Transactions in constraint handling rules.
\newblock In {\em Proc.\ of ICLP'08}. LNCS, vol. 5366. Springer, 516--530.

\bibitem[\protect\citeauthoryear{Sneyers, Schrijvers, and Demoen}{Sneyers
  et~al\mbox{.}}{2005}]{DBLP:conf/iclp/SneyersSD05}
{\sc Sneyers, J.}, {\sc Schrijvers, T.}, {\sc and} {\sc Demoen, B.} 2005.
\newblock Guard and continuation optimization for occurrence representations of
  chr.
\newblock In {\em Proc.\ of ICLP'05}. LNCS, vol. 3668. Springer-Verlag, 83--97.

\bibitem[\protect\citeauthoryear{Sneyers, {Van Weert}, Schrijvers, and {De
  Koninck}}{Sneyers et~al\mbox{.}}{}]{chr-survey}
{\sc Sneyers, J.}, {\sc {Van Weert}, P.}, {\sc Schrijvers, T.}, {\sc and} {\sc
  {De Koninck}, L.}
\newblock As time goes by: Constraint handling rules - a survey of chr research
  from 1998 to 2007.
\newblock To appear in TPLP.

\bibitem[\protect\citeauthoryear{Sulzmann and Lam}{Sulzmann and
  Lam}{2008}]{parallel-chr}
{\sc Sulzmann, M.} {\sc and} {\sc Lam, E. S.~L.} 2008.
\newblock Parallel execution of multi-set constraint rewrite rules.
\newblock In {\em Proc.\ of PPDP'08}. ACM Press, 20--31.

\end{thebibliography}
\end{document}